\documentclass[11pt, reqno, oneside]{amsart}
\newcommand{\col}{\color{black}}

 \usepackage{curves} 
\usepackage{graphicx} 
\usepackage{amsmath, amsfonts, amssymb, amscd,graphics, graphpap, tikz, color, xcolor, cancel,enumerate,enumitem, hyperref, mathabx, mathrsfs, mathtools}
\usepackage[scr=rsfs,cal=boondox]{mathalfa}

\usepackage{geometry}
\usetikzlibrary{arrows.meta,arrows}
\usetikzlibrary{decorations.pathreplacing,decorations.markings}
\tikzset{arrow data/.style 2 args={%
      decoration={%
         markings,
         mark=at position #1 with \arrow{#2}},
         postaction=decorate}
      }%
      
\usepackage[mathscr]{euscript}
  \usepackage[normalem]{ulem}
  \usepackage{soul}

\numberwithin{equation}{section}

\theoremstyle{plain}
\newtheorem{theo}{Theorem}[section]
\newtheorem{lem}[theo]{Lemma}
\newtheorem{prop}[theo]{Proposition}

\newtheorem{cor}[theo]{Corollary}

\theoremstyle{definition}
\newtheorem{rem}[theo]{Remark}

\newtheorem{example}[theo]{Example}
\newtheorem{definition}[theo]{Definition}

\newenvironment{pf}{\noindent{\it Proof. }}{$\hfill\square$\par\medskip}

\theoremstyle{plain}

\theoremstyle{definition}

\newcommand{\beq}{\begin{equation}}
\newcommand{\eeq}{\end{equation}}
\newcommand{\beqn}{\begin{equation*}}
\newcommand{\eeqn}{\end{equation*}}
\renewcommand{\a}{\alpha}
\renewcommand{\b}{\beta}

\renewcommand{\d}{\delta}
\newcommand{\e}{\epsilon}
\newcommand{\ve}{\varepsilon}
\newcommand{\f}{\varphi}
\newcommand{\g}{\gamma}
\newcommand{\h}{\eta}

\renewcommand{\k}{\kappa}

\renewcommand{\o}{\omega}
\newcommand{\q}{\vartheta}

\newcommand{\s}{\sigma}

\newcommand{\D}{\Delta}

\newcommand{\F}{\Phi}
\newcommand{\G}{\Gamma}

\renewcommand{\L}{\Lambda}

\newcommand{\bC}{\mathbb{C}}
\newcommand{\bD}{\mathbb{D}}
\newcommand{\bR}{\mathbb{R}}

\newcommand{\gd}{\mathfrak{d}}

\newcommand{\gr}{\mathfrak{r}}
\newcommand{\gs}{\mathfrak{s}}

\newcommand{\gB}{\mathfrak{B}}

\newcommand{\cC}{\mathcal{C}}
\newcommand{\cD}{\mathscr{D}}

\newcommand{\cH}{\mathscr{H}}

\newcommand{\cK}{\mathscr{K}}
\newcommand{\cL}{\mathscr{L}}
\newcommand{\cM}{\mathscr{M}}

\newcommand{\cP}{\mathscr{P}}
\newcommand{\cQ}{\mathscr{Q}}

\newcommand{\cS}{\mathscr{S}}

\newcommand{\cU}{\mathscr{U}}
\newcommand{\cV}{\mathscr{V}}
\newcommand{\cW}{\mathscr{W}}

\newcommand{\p}{\partial}

\renewcommand{\square}{\kern1pt\vbox
{\hrule height 0.6pt\hbox{\vrule width 0.6pt\hskip 3pt
\vbox{\vskip 6pt}\hskip 3pt\vrule width 0.6pt}\hrule height0.6pt}\kern1pt}

\DeclareMathOperator\sign{sign}

\DeclareMathOperator\Id{Id}

\renewcommand\Re{\operatorname{Re}}
\renewcommand\Im{\operatorname{Im}}

\renewcommand\={:=}

\newcommand{\wt}{\widetilde}
\newcommand{\wh}{\widehat}

\newcommand{\bt}{\begin{theo}}
\newcommand{\et}{\end{theo}}
\newcommand{\bp}{\begin{prop}}
\newcommand{\ep}{\end{prop}}
\newcommand{\bc}{\begin{cor}\ \ }
\newcommand{\ec}{\end{cor}}
\newcommand{\bl}{\begin{lem}\ \ }
\newcommand{\el}{\end{lem}}
\newcommand{\bd}{\begin{definition}}
\newcommand{\ed}{\end{definition}}
\newcommand{\n}{\nabla}

\newcommand{\be}{\begin{equation}}
\newcommand{\ee}{\end{equation}}

\def\<#1,#2>{\langle\,#1,\,#2\,\rangle}
\newcommand{\arr}{\begin{array}{rlll}}
\newcommand{\ea}{\end{array}}
\newcommand{\bea}{\begin{eqnarray}}
\newcommand{\eea}{\end{eqnarray}}
\newcommand{\bean}{\begin{eqnarray*}}
\newcommand{\eean}{\end{eqnarray*}}

\newcommand\wc{\widecheck}

\newcommand{\GGa}[3]{{\mathbf \G}_{#1 #2}^{\phantom{#1} #3}}

\font\smallsmc = cmcsc9
\font\smalltt = cmtt8
\font\smallit = cmti8

\def\sideremark#1{\ifvmode\leavevmode\fi\vadjust{
\vbox to0pt{\hbox to 0pt{\hskip\hsize\hskip1em
\vbox{\hsize2cm\tiny\raggedright\pretolerance10000
\noindent #1\hfill}\hss}\vbox to8pt{\vfil}\vss}}}

\renewcommand{\sf}{shearfree }
\newcommand{\I}{\operatorname{i}}

\renewcommand{\e}{\operatorname{e}}

\renewcommand{\Re}{\operatorname{Re}}
\renewcommand{\Im}{\operatorname{Im}}

\newcommand{\ps}{{\operatorname{p}}}
\newcommand{\qs}{{\operatorname{q}}}

\newcommand{\cs}{\operatorname{\mathsf{b}}}

\newcommand{\Ric}{{\operatorname{Ric}}}
\newcommand{\Scal}{{\operatorname{Scal}}}
\newcommand\bex{\begin{example}}
\newcommand\eex{\end{example}}
\newcommand\br{\begin{rem}}
\newcommand\er{\end{rem}}

\newcommand{\bfB}{{\mathsf B}}
\newcommand{\bfa}{{\mathsf m}}
\newcommand{\bfm}{\wt {\mathsf m}}
\newcommand{\bff}{{\mathsf f}}

\newcommand{\bfA}{{\mathsf A}}
\let\underbrace\LaTeXunderbrace

\makeatletter
\@namedef{subjclassname@1991}{2020 Mathematics Subject Classification}
\makeatother

\title[Null fluid gravitational fields on Kerr manifolds]{Null fluid gravitational fields on Kerr manifolds\\ and optical    lifts of Sasaki  structures}
 \author[Masoud Ganji, Cristina Giannotti and  Andrea Spiro]{Masoud Ganji $\dagger \ddagger$ \quad Cristina Giannotti ${}^*$  \quad  Andrea Spiro${}^\circ$}
\date{}

\begin{document}
 \thanks{{\it Data availability statement.} 
No datasets were generated or analysed during the current study.}
\begin{abstract}
 Building on the characterisation in  [C. D. Hill, J. Lewandowski and P. Nurowski, 
 Indiana Univ. Math. J.
  {\bf 57}  (2008),  3131--3176]  of $4$-dimensional Lorentzian metrics adapted to an optical structure and satisfying the null fluid Einstein equations, we give an explicit parameterisation of this class under the assumption that the optical structure is of Kerr type. As immediate consequences, we obtain: (1) a new method for constructing solutions to the Einstein equations with  null fluid energy momentum tensor, yielding  a large family of explicit metrics that naturally includes the classical Kerr black hole metrics and all Ricci flat examples described in [M. Ganji, C. Giannotti, G. Schmalz and A. Spiro, Ann. Physics 
{\bf 75} (2025), Paper No. 169908, 28]; (2) a  solution to a conjecture in  Hill, Lewandowski and Nurowski's paper  on the local existence of smooth optical lifts in the case of Sasaki CR structures. 
 \par
 \par
 \vskip 1 cm 
 \font\smallsmc = cmcsc8
\font\smalltt = cmtt8
\font\smallit = cmti8
\hbox{\parindent=1.5cm \parskip=0pt
\vbox{\baselineskip 9.5 pt \hsize=3.7truein
\obeylines
{\smallsmc 
$\dagger$  School of Science and Technology
University of New England,
Armidale NSW 2351
Australia
}\medskip
{\smallit E-mail}\/: {\smalltt mganjia2@une.edu.au
\ \\
\ \\
$\ddagger$=corresponding author
\ \\
}
}
\hskip - 3.0truecm
\vbox{\baselineskip 9.5 pt \hsize=3.7truein
\obeylines
{\smallsmc
${}^*, {}^\circ$ Scuola di Scienze e Tecnologie
Universit\`a di Camerino
Via Madonna delle Carceri
I-62032 Camerino (Macerata)
ITALY
\ 
}\medskip
{\smallit E-mail}\/: {\smalltt cristina.giannotti@unicam.it
\smallit E-mail}\/: {\smalltt andrea.spiro@unicam.it}
\ \\
\ \\
}
}
\end{abstract}

\maketitle
\ \\
\newpage
\section{Introduction} 
We recall that an {\it  optical structure}
on an $n$-dimensional  manifold  $M$ is  a pair $\cQ = (\cK, \{g\})$ given by
\begin{itemize}[leftmargin=15pt]
\item[--] a $1$-dimensional distribution $\cK \subset TM$,  
\item[--] the unique family   $\{g\}$  of  all Lorentzian metrics $g$,   such that: (1) the  curves  tangent to  $\cK$ determine   a   geodesic \sf null congruence on $M$;  (2) the $g$-orthogonal distribution $\cK^{\perp_g}$ is a fixed distribution $\cW\subset TM$, independent of $g\in\{g\}$ (so $\cW=\cK^{\perp_g}$ for all $g$). 
\end{itemize} 
The metrics in the class  $\{g\}$  are said to be    {\it compatible}   with the optical structure.  \par
\smallskip
The notion of optical structure  was  introduced by Robinson and Trautman  to encode the geometric properties  of   Lorentzian manifolds  admitting   electromagnetic plane waves (or   appropriate higher dimensional generalisations)  propagating  along  a prescribed foliation by   null  geodesics  (see e.g. \cite{RT, RT1, RT11, FLT, AGS, AGSS,  Tafel1, Rob, NT} and references therein). 
A standard way  to  construct  explicit examples of  optical structures is  as follows.\par
\medskip
 Consider a  $(2k-1)$-dimensional manifold $\cS$,  equipped  with a $(2k-2)$-dimensional contact distribution $\cD$ and with a field $J$ of complex structures $J_x: \cD_x \to  \cD_x$ on the spaces of the distribution $\cD$  satisfying, for any  vector fields  $X, Y\in\cD$,  
$$[JX, JY] - [X, Y] \in \cD \ ,\qquad [X,Y]-[JX,JY]+J([JX,Y]+[X,JY]) = 0\ .$$
A triple $(\cS, \cD, J)$ of this kind is  called  {\it Levi non-degenerate CR manifold}. If  there exists a contact form $\theta$ for  $(\cS, \cD)$ (i.e., a $1$-form with $\ker \theta_x  = \cD_x$, $x \in M$)  such that   $d \theta_x( \cdot, {\col J_x} \cdot)$ is a positive definite  scalar product on each  $\cD_x \subset T_x M$, then $(\cS, \cD, J)$ is called {\it strongly pseudoconvex}. \par
\smallskip
Consider  the  trivial  principal $\bR$-bundle $\pi^{\cS}: M = \cS \times \bR \to \cS$ over a strongly pseudoconvex CR manifold $(\cS, \cD = \ker \theta, J)$. Denote by  $\q$ and $\ps_o$    the   pulled-back $1$-form   $\q = \pi^{\cS*} \theta$  on $M$ and the vertical vector field  $\ps_o = \frac{\p}{\p u}$  generating the  natural  right $\bR$-action on $M = \cS \times \bR$, respectively.    Then  let  $\cQ^{(\cS)} = (\cK, \{g\})$ be the pair given by the  $1$-dimensional distribution $\cK = \langle \ps_o \rangle \subset T M$  and the class $\{g\}$ of  Lorentzian metrics on $M$  having the form
$$ g = \s \pi^{\cS*}(d \theta( \cdot, {\col J}\cdot) )+ \q \vee \bigg(\wt \a \ps_o^* + \gd +  \frac{\wt \b}{2} \q\bigg)\ ,\qquad \text{where}$$
\begin{itemize}[leftmargin = 20 pt]
\item[(i)]  $\ps_o^*$  is an  $\bR$-invariant $1$-form with  $ \ps_o^*(\ps_o) = 1$  and  $\cH \= \ker \ps_o^* \cap \ker \q$  is the unique $\bR$-invariant  distribution in $TM$ that projects onto $\cD \subset T\cS$, 
\item[(ii)]  $\s, \wt \a, \wt \b$  are freely specifiable smooth  functions with  $\s, \wt \a$ nowhere vanishing, 
\item[(iii)] $\gd$ is a freely specifiable $1$-form with $\ker \gd$ containing  a distribution  $\cV \subset TM$ which is complementary to $\langle \ps_o\rangle + \cD$. 
\end{itemize}
One checks directly  that any such  $\cQ^{(\cS)} = (\cK, \{g\})$    is an optical structure (\cite{RT11, HLN, AGSS, AGSS1, FLT}). These optical structures provide an important   link between  the  geometry of the strongly pseudoconvex CR manifolds $(\cS, \cD, J)$ and Lorentzian metrics $g$  compatible with the corresponding optical structures $\cQ^{(\cS)}$.   Indeed, such a relation  led to very interesting results in both  CR and Lorentzian geometry; see  for instance \cite{HLN}  for a number of remarkable 
results and applications  in both areas. \par
\medskip

In this paper we deal with the optical structures built as above from a $3$-dimensional strongly pseudoconvex CR manifold $(\cS, \cD = \ker \theta, J)$ under the assumption that: (a) $\cS$ is a principal $\bR$-bundle $\pi: \cS \to N$ over  a   $2$-dimensional K\"ahler  manifold  $(N, J, g_o)$; (b) the distribution $\cD$ is  $\bR$-invariant and complementary to the vertical distribution of $\cS$; (c) the contact $1$-form $\theta$ is   a connection $1$-form $\theta: T \cS \to \bR$   satisfying  $d \theta = \pi^*\o_o$, where  $\o_o \= g_o(J \cdot, \cdot)$  is the K\"ahler form of $(N, J, g_o)$.  We mention that the   CR structures satisfying (a) -- (c) belong to  the  class of the so called  {\it Sasaki structures}.  We call   the  optical structures built upon these Sasaki manifolds  {\it of Kerr type} and we call  any $4$-manifold $M \simeq  \cS \times \bR$,  equipped with such an optical structure,  a {\it Kerr manifold}. These  names are motivated by the fact that  the classical   Kerr metrics of rotating black holes  are   compatible metrics of   Kerr type optical structures (see \cite{GGSS} and references therein for details). \par
\medskip
Our  first main result can be summarised as follows.  In \cite{HLN}  Hill, Lewandowski and Nurowski  determined a  local characterisation of  the $4$-dimensional  Lorentzian metrics  $g$ that are  
compatible with an optical structure determined as above  by an {\it embeddable} strongly pseudoconvex CR manifold (i.e. equivalent to a strongly pseudoconvex real hypersurface in $\bC^2$) and satisfy  an equation of the form 
\beq \label{1}  \Ric = \L g + \bff \q \vee \q\qquad \text{for some function} \ \bff\ .\eeq  
Note that,  for any such a compatible metric $g$, the tensor field $\q \vee \q$ is null and  \eqref{1} is the Einstein equation  for the gravitational field  of a null fluid in a background metric with cosmological constant $\L$.  In addition, by  the generalisation  in  \cite[Thm. 5.10]{GHN} of the Goldberg-Sachs Theorem, it is known that the curvature  of any such compatible metric is algebraically special  at all points. Following \cite{GSS, SKH}, we call metrics of this kind {\it quasi-Einstein}. \par
The results in \cite{HLN} on quasi-Einstein metrics  are presented using complex functions in  a special system  of  {\it complex} coordinates $(z, w)$ on $\bC^2$ and  satisfying a particular set of differential  constraints.  Subsequently,  in \cite{GSS} the authors studied  the  quasi-Einstein metrics  compatible with  Kerr type optical structures, a class to which     Hill, Lewandowski, and Nurowski's  results immediately apply with  substantial simplifications.  In this paper, we further develop  \cite{HLN, GSS} to provide a fully explicit and coordinate-free  local parametrisation  of    quasi-Einstein compatible metrics on a  $4$-dimensional Kerr manifold. We present them in terms of eight  functions {\col or, more precisely, a constant (= the freely specifiable  constant $\L$) and seven real functions}   defined on the K\"ahler surface,   onto which the Kerr manifold projects,  constrained by a system of five   partial differential equations,  the first four readily solvable. 
\par
\smallskip
This  result has at least two  remarkable   consequences.  First, it  simplifies  the  presentation  of the quasi-Einstein compatible metrics on a Kerr manifold so greatly that  we can resolve, in a very simple way,  the following conjecture by Hill, Lewandowski and Nurowski in the setting of Kerr manifolds. In \cite{HLN} it was proved that {\it for any choice of the cosmological constant $\L$ and for any   real analytic and embeddable  strongly pseudoconvex $3$-manifold $(\cS, \cD, J)$, there locally exists at least one quasi-Einstein Lorentzian metric   with curvature tensors of Petrov type II or D, which is compatible with the optical structure on $M = \cS \times \bR$  described above. }  We call these compatible metrics {\it  (local) optical lifts} of  the CR structure $(\cD, J)$.
 In  \cite{HLN}  the authors conjectured   that the real analyticity  is not essential  for the above result and that it  holds under weaker regularity  hypotheses.  Using results on logistic elliptic equations, this was partially confirmed in \cite{GSS} for Kerr type optical structures, under the  additional assumptions that  the cosmological constant $\L$ is positive and that the underlying  K\"ahler surface $N$ is biholomorphic to $\bC$.  
 Exploiting our new parametrisation of quasi-Einstein metrics on Kerr manifolds, we now show that, for Kerr type optical structures, the conjecture holds with no restriction on the sign of $\L$  and on the K\"ahler surface $N$.  In fact,  this  follows directly from a classical general result on non‑linear elliptic equations.\par 
\smallskip

The second consequence of our first result is that,  {\col for any choice of  the  constant $\L$,} by setting two of the {\col seven} parametrising functions  equal to zero and assuming two of the remaining functions  proportional to each other, we obtain a family of quasi-Einstein metrics effectively parametrised by: (a) three freely specifiable constants, (b) one freely specifiable harmonic function {\col vanishing at the origin}, and (c) a real function of two variables subject to a single non-linear elliptic equation. This family contains the classical Kerr metrics and the large class of explicit Ricci flat Lorentzian metrics recently constructed in \cite{GGSS}; we therefore refer to it as the \emph{enlarged Kerr family}. 
 Using arguments from \cite{GGSS}, we construct a  large subclass of  quasi-Einstein metrics within the enlarged Kerr family. These furnish an infinite family of pairwise non isometric metrics representing gravitational fields sourced by non-vanishing null fluid energy momentum tensors, which, to the best of our knowledge, have not previously appeared in the literature. 
  Note also that, as a consequence of the  cited curvature properties  of quasi-Einstein metrics  and the fact that  the Weyl scalar $\Psi_2$  with respect to the  tetrads   considered in \cite{HLN, ONeill},  is  generically  non-zero,  the enlarged Kerr family provides a broad  family of metrics whose  curvature tensors are of Petrov  type II or D. \par 
\smallskip
The paper is organised as follows. Preliminaries are presented in \S2 and \S3. In \S 4 we establish our first main result,  i.e., a new parametrisation of quasi-Einstein metrics compatible with Kerr  type optical structures. In \S 5 we introduce and analyse the above mentioned enlarged Kerr family of quasi-Einstein metrics. In \S 6 we prove the announced resolution of the conjecture in \cite{HLN} on the local existence of $\cC^\infty$   quasi-Einstein optical lifts  in the context of Kerr type optical structures.\par
\medskip
\noindent{\it  Acknowledgement.} We warmly thank Marcello Ortaggio for  helpful critics and suggestions.\par
\bigskip
\section{Preliminaries (I): Optical geometries and optical lifts}
\subsection{Geodesic shearfree null congruences and optical geometries}
\label{section1}
Let $(M, g)$ be  a  Lorentzian $n$-manifold.  A {\it  null congruence on $M$}  is a   foliation  by   null curves.  Let $\{\ps\}$ be the class of null vector fields tangent to the curves of such a null congruence, $ \cW$   the $g$-orthogonal distribution to  all  vector fields   in $\{\ps\}$, and  $ h= g|_{\cW\times \cW}$  the degenerate metric induced  on $\cW$.
 The null congruence is   called {\it \sf} if  there  exists a vector field  $\ps$ of the class  $ \{\ps\}$, whose  (local)  flow  preserves  the distribution $\cW$ and the conformal class  $[h]$ of  the degenerate metric $h$ on $\cW$.
%
It is known that this condition is    equivalent to   (see e.g. \cite[Lemma 2.1]{AGSS})  
\beq \label{defsf2}   \cL_{\ps} g = f g + {\ps}^\flat \vee \h\qquad \text{for some function $f$ and a $1$-form}\ \ \h\ \eeq
and that,  if this  holds,  all curves of the null congruence are geodesics (\cite[Prop. 2.4]{AGSS}). Due to this, in Physics, a  foliation of this kind is usually called   {\it geodesic \sf null congruence}.
\par
\smallskip
An {\it optical geometry} on a manifold $M$ is a  quadruple 
$\cQ = (\cW, [h]_{\pm}, \cK, \{g\})$, 
 given by 
\begin{itemize}[leftmargin = 20pt] 
\item[(i)] a codimension one distribution $\cW \subset TM$,  
\item[(ii)] the union  $[h]_{\pm} = [h] \cup [-h]$  of  two conformal classes   $[h]$, $[-h]$,  of    degenerate   metrics   on  $\cW$ with  one-dimensional kernels, the first consisting of semipositive metrics,  the second made of their opposites, 
\item[(iii)] the $1$-dimensional distribution $\cK$ defined by  $\cK|_x = \ker h_x \subset \cW_x$, $x \in M$, $h \in [h]_\pm$, 
\item[(iv)]  the class $\{g\}$ of all  Lorentzian metrics  (some with   mostly plus, others with  mostly minus signature)   that induce on $\cW$ at least one of the   degenerate metrics in  $ [h]_{\pm}$, 
\end{itemize}
such that the following holds:  {\it the class $\{g\}$ is not empty and  there exists a vector field  $\ps$ in   $\cK$ satisfying  \eqref{defsf2}}.
 The  metrics of the class  $\{g\}$ of the optical structure $\cQ$ are said  to be {\it compatible with $\cQ$}.\par
 \smallskip
We  recall that  for   any optical geometry  $\cQ = (\cW, [h]_{\pm}, \cK, \{g\})$,   either one of the  pairs   $(\cW, [h]_{\pm})$  or $( \cK_h = \ker h, \{g\})$  is   sufficient  to  recover the other  terms of the quadruple.  The pairs  $(\cW, [h]_{\pm})$  that appear in  optical geometries  are called    {\it \sf structures} in  \cite{AGSS},   while   the corresponding  pairs $(\cK, \{g\})$ are precisely  the    {\it optical geometries} as they were defined by   Robinson and Trautman  in  \cite{RT} (provided that  both definitions are relaxed to allow  mostly plus and mostly minus signatures, of course). 
\par
\smallskip
Given  an optical geometry $\cQ$, all  metrics  that are compatible with $\cQ$ determine the same   \sf  null congruence, i.e., the foliation by  the integral curves of $\cK$. For this reason, an optical geometry can be regarded as the  geometry underlying a prescribed \sf  null congruence.
 \par 
\smallskip
\subsection{Robinson and Trautman  optical geometries    and CR structures}
 Consider now an optical geometry $\cQ  = (\cW, [h]_{\pm}, \cK, \{g\})$ on a manifold $M$,  where  the $1$-dimensional distribution $\cK$  is generated by a vector field $\ps$, which   not only preserves the codimension one distribution $\cW$, but is  also complete and  generates a  group   $A = \{e^{t{\ps}}, t \in \bR\}$ of diffeomorphisms (isomorphic  to $\bR$ or $S^1$)  acting freely and properly.  This is equivalent to saying that $\cQ$ is an optical geometry on the total space of the principal  $A$-bundle ${\col \pi^\cS}: M \to \cS = M/A$, with the distribution $\cK$ equal to the vertical distribution $\cK \subset T^{\operatorname{vert}} M$ of the  bundle.  Following the terminology introduced in \cite{AGSS}, we call  the optical geometries of this kind {\it  Robinson and Trautman geometries} (or just  {\it RT  geometries}, for short). \par
 \medskip
 \begin{rem} \label{rem1} Notice that {\it any optical geometry is \underline{locally} identifiable with an RT geometry}. Indeed,  if $\wt \ps$  is a vector field on a manifold $M$, that is tangent to a \sf congruence and  whose local flow preserves the distribution $\cW$ and the conformal class $[h]$  of  an optical geometry $\cQ$, then around any  $x \in M$, we may choose  coordinates $(x^0, x^1, \ldots x^{n-1})$,  in which the flow of $\wt \ps$ takes the translational form $\Phi^{\wt \ps}_t( x^0, x^1, \ldots, x^{n-1}) = (x^0 + t, x^1, \ldots,  x^n)$. In this way, we may determine a local diffeomorphism between a neighbourhood of $x \in M$ and an open set of a trivial bundle of the form ${\col \pi'}: M' =  \cS \times A \to  \cS$  with $A = S^1$ or $A = \bR$, mapping $\wt \ps$ into the vector field $\ps = \frac{\p}{\p r}$ whose flow generates the standard translations  of the $1$-dimensional group  $A$.  One can directly check that under such a local diffeomorphism, the optical geometry $\cQ$ is locally mapped into an RT geometry  $\cQ'$ on 
 $M' =  \cS\times A$.  Due to this observation, {\it if one is interested only in {\rm local} properties,  there is essentially no loss of generality in  restricting to   the class of RT geometries}.  
 \end{rem} 
Each  RT  geometry on the total space of a bundle   ${\col \pi^\cS}: M \to \cS = M/A$ is known to be  determined by a uniquely associated   {\it sub-conformal structure} on the base space  $\cS$, i.e.,   by a  codimension one distribution $\cD$   equipped   with  an  equivalence class $[g^\cD]_{\pm}$  of Riemannian or negative definite metrics  on $\cD$, each determined from the others by multiplication with a nowhere-vanishing conformal factor. In fact, 
\begin{prop}[{\cite[Prop. 3.6]{AGSS}}] \label{propino} Given an $A$-bundle  ${\col \pi^\cS}: M \to \cS $, with $A = \bR$ or $S^1$, 
there is a natural one-to-one correspondence between  the  RT  geometries  $\cQ  = (\cW, [h]_{\pm}, \cK, \{g\})$ on $M$  with $\cK = T^{\operatorname{vert}}M$ and the  sub-conformal structures $(\cD, [g^\cD]_{\pm})$ on  $\cS$.
\end{prop}
A distinguished class of RT  geometries is given by the   even-dimensional total spaces of $A$-bundles $M$,  associated with   sub-conformal structures $(\cD, [g^\cD])$  on their base spaces, where    $\cD \subset T (M/A)$ is a contact distribution. These  are called     {\it (maximally) twisting} RT geometries.   They are particularly interesting   because  there exists a  natural correspondence between them and    an appropriate class of   almost  CR structures. Let us briefly review this correspondence.  \par
\smallskip
Let $\cS$ be  an odd-dimensional  manifold  with $\dim \cS = 2n -1$. A  {\it (codimension one) almost CR structure on $\cS$}  is a  pair   $(\cD, J)$,  consisting of: 
\begin{itemize}[leftmargin = 18pt]
\item[--]  a codimension-one distribution  $\cD \subset T\cS$, and 
\item[--]  a  smooth  family  $J$ of endomorphisms of the  subspaces $\cD_x \subset T_x \cS$, $x \in \cS$,  such that  $J^2_x =  - \Id_{\cD_x}$ for all $x$.
\end{itemize}
  An almost  CR structure  on $\cS$ is called {\it integrable} if the following two conditions hold:
\begin{itemize}[leftmargin = 18pt]
\item[(1)] For  any pair of vector fields $X, Y$ in $\cD$, the vector field  $[X, J Y] + [JX, Y]$ also lies in $\cD$; 
\item[(2)] The Nijenhuis tensor field $N_J \in \cD^* \otimes \cD^* \otimes \cD$ of $J$,  defined by
$$N_J(X,Y)=[X,Y]-[JX,JY]+J([JX,Y]+[X,JY]) \ , \ \ X, Y \in \cD ,$$
vanishes identically. 
\end{itemize}
A {\it CR structure} is an  integrable almost CR structure.  If  an almost CR structure satisfies only condition  (1)  (but not necessarily (2)),   it  is  called   {\it partially integrable}.\par
\medskip
Note that if   $\dim \cS = 3$, then  {any almost CR structure on $\cS$ is  integrable}.\par
 \medskip
 Let $(\cD, J)$ be a partially integrable  CR structure on $\cS$ and let $\theta$ be a {\it defining $1$-form  for $\cD$}, i.e., a $1$-form   such that $\ker \theta_x = \cD_x$ for every $x \in \cS$. One can directly check that, in this case,   the type $(0, 2)$ tensor field  on $\cD$ defined by 
$$h^\theta \in \cD^* \times \cD^*\ ,\qquad h^\theta(X, Y)\= d \theta(X, JY)$$
  is  symmetric.  This is called the  {\it Levi form}  of the  partially integrable CR structure, and  it is independent of  the  choice of   $\theta$ {\it up to a multiplication by a  nowhere vanishing conformal factor}. Note that when   $\cD$ is   contact,  the  Levi form  $h^\theta = d \theta(\cdot, J\cdot)$ is non-degenerate.  If it  is positive definite, the  partially integrable CR structure is said to be  {\it strongly pseudoconvex}.  
\par
\smallskip
We are now ready to  state  the precise  relation between twisting RT  geometries and CR structures. The following   is a direct application  of  \cite[Cor. 4.5]{AGSS}. 
\begin{theo}  \label{cor35} Let $\pi: M \to \cS = M/A$ be a principal bundle with $1$-dimensional structure group $A = \bR$ or $S^1$, and with even-dimensional total space $M$. Then there are natural (with respect to  $A$-bundle automorphisms)  bijections between the following three classes: 
\begin{itemize}[leftmargin = 18pt]
\item[(1)] The  twisting RT  geometries   $\cQ  = (\cW, [h]_{\pm}, \cK , \{g\})$ on $M$ with $\cK =   T^{\operatorname{vert}}M$; 
\item[(2)] The sub-conformal structures $(\cD, [g^\cD]_{\pm})$ on   $\cS$   where $\cD$ is a contact distribution; 
\item[(3)] The pairs  $((\cD, J), [B]_{\pm})$,  consisting of 
\begin{itemize}[leftmargin = 10pt]
\item[--] a  strongly pseudoconvex partially integrable  CR structure $(\cD, J)$; 
\item[--] a  class  $[B]_{\pm}$ of  smooth fields of endomorphisms $B_x: \cD_x \to \cD_x$, $x \in M$, each  defined up to multiplication  by a nowhere vanishing factor, and either   positive definite or negative definite with respect to the positive Levi form of $(\cD, J)$.
\end{itemize}
\end{itemize}
\end{theo}
\noindent Explicit  expressions for the bijections  between the classes (1) -- (3) in Theorem \ref{cor35},  given in full detail, can be found  in \cite{AGSS}.
\par
\smallskip
\subsection{Optical  lifts of   Sasaki  CR  structures} 
\label{optquant}
 An immediate  and important consequence of Theorem \ref{cor35} is that the twisting RT geometries on a  fixed  $2n$-dimensional  principal $A$-bundle ${\col \pi^\cS}: M \to \cS$, $A = \bR$ or $S^1$, are naturally grouped according to  the corresponding partially integrable     CR structures on $\cS$. More precisely,   two RT geometries  $$\cQ  = (\cW, [h]_{\pm}, \cK , \{g\}) \ ,\qquad \cQ'  = (\cW', [h']_{\pm}, \cK' = \cK , \{g'\})$$ on 
the   $A$-bundle $\pi: M \to \cS$ are said to  {\it have  the same underlying CR structure} if the corresponding  pairs  $((\cD, J), [B]_{\pm})$ and $((\cD', J'), [B']_{\pm})$ in Theorem \ref{cor35} (3) satisfy
$$(\cD, J) = (\cD', J')\ .$$
Due to the naturality of the bijection  given in  Theorem \ref{cor35}, if   $(\cD, J)$ is a  strongly pseudoconvex partially integrable CR structure on $\cS$, and   $\{\{g\}\}$ denotes  the  class of  Lorentzian metrics on $M$ compatible  with  at least one  RT geometry having  $(\cD, J)$ as underlying CR structure, then any bundle   automorphism of ${\col \pi^\cS}: M \to \cS$,  that projects onto an  automorphism of the CR structure   $(\cD, J)$ maps  the  class $\{\{g\}\}$ into itself. Conversely, any  automorphism of the CR structure   $(\cD, J)$ that lifts to a bundle automorphism of ${\col \pi^\cS}: M \to \cS$, preserves the class $\{\{g\}\}$.  
 \par
 \medskip
 Following the terminology adopted  in \cite{HLN, GSS}, throughout this paper we  refer to the metrics    in  such invariant class  $\{\{g\}\}$ as the   {\it optical (Lorentzian) lifts} of the  almost CR structure $(\cD, J)$.  By  the above discussion, any  invariant of  the optical   lifts  {\it considered  determined  up to the action of a bundle  diffeomorphism preserving the entire class $\{\{g\}\}$}, canonically corresponds  to   an invariant of the  underlying  almost  CR structure $(\cD, J)$. \par
 \medskip
 Let $\cS$ be a $(2n-1)$-dimensional manifold  equipped with  a  strongly pseudoconvex {\it integrable}  CR structure $(\cD ,J)$.   The triple $(\cS, \cD, J)$ is called a {\it regular Sasaki manifold}
if  it is the total space of a  principal  $A'$-bundle $\pi: \cS \to N$, with $A' = \bR$ or $S^1$,  such that the right  action of   $A'$ leaves  the CR structure invariant.  \par
\medskip 
If $(\cS, \cD, J)$ is a regular Sasaki manifold, then:
\begin{itemize}[leftmargin = 18 pt]
\item[(a)]  There exists an $A'$-invariant defining  $1$-form $\theta$ for $ \cD$, which also serves as  a connection $1$-form for  the $A'$-bundle;  
\item[(b)]  The complex structure $J$ and the Levi form $h^\theta$ of $(\cD, J)$ project onto an integrable complex structure $J$ and a Riemannian metric $g_o$ on $N$,  making $(N, g_o, J)$ a K\"ahler manifold; 
\item[(c)] The K\"ahler form $\o_o \= g_o(J\cdot,\cdot )$  of $g_o$  satisfies $d\theta= \pi^* \o_o$. 
\end{itemize}
\par
\smallskip
Conversely, 
a K\"ahler manifold $(N, J, g_o)$  with K\"ahler form $\o_o$  is called {\it quantisable} if there exists a principal $A'$-bundle $\pi: \cS \to N$,  $A' = S^1$ or $\mathbb{R}$,  equipped with
a connection $1$-form $\theta: T\cS \to \bR$ satisfying the condition 
$d\theta= \pi^* \o_o$.
In this case, the  horizontal distribution $\cD = \ker \theta$ on $\cS$ is contact, and the lift of the 
complex structure  $J$ on  $N$   to $\cD$  makes $(\cS, \cD, J)$ a regular Sasaki manifold.
\par
\smallskip
We stress the fact that {\it the correspondence between regular Sasaki manifolds and quantisable K\"ahler manifolds described above is a bijection} (see \cite{ACHK}).
\par
\smallskip
Consider now a  regular Sasaki $(2n - 1)$-manifold $(\cS, \cD = \ker \theta,J)$ and denote by  $(N, J, g_o)$ its associated  quantisable  K\"ahler $(2n-2)$-manifold. Let also    $\pi^{\cS}: M = \cS \times \bR \to \cS$   be the  trivial   $\bR$-bundle over $\cS$, and let    $\q$ and $\ps_o$ denote, respectively,   the   pulled-back $1$-form   $\q = \pi^{\cS*} \theta$   and the vertical vector field  $\ps_o = \frac{\p}{\p r}$  generating the  $\bR$-action on $M = \cS \times \bR$.   According to Theorem \ref{cor35} there is a one-to-one correspondence between the twisting RT geometries on $M = \cS \times \bR$, with $1$-dimensional distribution $\cK_o = \langle \ps_o \rangle  = T^{\operatorname{vert}} M \subset T M$ and the pairs $((\cD, J), [B]_{\pm})$ as in item (3) of that theorem, where  $(\cD, J)$  is  the CR structure on $\cD$.  We recall that, following the  proof of   \cite[Cor. 4.5]{AGSS}, all such 
RT geometries can be explicitly determined. \par
\medskip
Among  them, the simplest RT geometry  is the one  corresponding to the class of endomorphisms  $[B = \Id_\cD]_{\pm}$. Such RT geometry turns out to be  the  quadruple  $\cQ_o = (\cW_o, [h_o]_{\pm}, \cK_o  = \langle \ps_o \rangle, \{g\})$, defined as follows (for further details,  see e.g., \cite{AGSS, AGSS1, GGSS}):  
\begin{itemize}[leftmargin = 15pt]
\item   $\{g\}$ is the set of the Lorentzian metrics on $M$  having the form
\beq\label{comp1}  g =\s \wc \pi^* g_o + \q \vee \bigg(\s \a \ps^*_o + \s \gd +   \frac{\s \b}{2} \q\bigg), \eeq
where we use the following  notation:
\begin{itemize}[leftmargin = 20 pt]
\item[(i)] $ \wc \pi \= \pi \circ \pi^{\cS} $; 
\item[(ii)]  $\ps_o^*$   is  the unique $\bR$-invariant $1$-form, satisfying the condition $ \ps_o^*(\ps_o) = 1$, with   $\cH \= \ker \ps_o^* \cap \ker \q$  equal to  the unique $\bR$-invariant horizontal distribution  projecting onto $\cD \subset T \cS$, 
\item[(iii)]  $\s, \a, \b$  are  freely specifiable $\cC^\infty$  functions with  $\s, \a \neq 0$ at all points, and   $\gd$ is a  freely specifiable $1$-form with the property that $\ker \gd$ contains the vertical distribution of $\wc \pi : M \to N$, 
\end{itemize}
\item $\cW$ is the   $g$-orthogonal distribution $\cW  = \cK^{\perp_g}$  for one (and hence, for any) metric  $g \in \{g\}$; 
\item $[h_o]_{\pm}$  is the pair   of conformal classes of the degenerate metrics $h =  \pm g|_{\cW \times \cW}$ with  $g \in \{g\}$. 
\end{itemize}
 We call  this  the {\it standard RT geometry of the $A \times \bR$-bundle $\wc\pi = \pi^\cS \circ \pi: M = \cS \times \bR \to N$}.
Note that each compatible metric $g$ of this optical geometry is  completely determined by the quadruple  $(\s, \a, \b, \gd)$,  consisting of three   real functions $\s$, $\a$, $\b$, with $\s$, $\a$ nowhere vanishing,  and  the tensor field $\gd \in \cH^*$.
We call this quadruple the {\it (canonical) datum} of  $g$.\par
\medskip

\section{Preliminaries (II): Null fluid gravitational fields on Kerr manifolds }
 \subsection{An  important  class of optical  geometries:  the Kerr structures} \label{sect2} Following the same line of reasoning as in Remark \ref{rem1},  we point out   that  the total space of a  principal bundle $\pi^\cS: M \to \cS $ with a $1$-dimensional  structure group  $A = \bR$ or $S^1$,  is always locally diffeomorphic  to a  trivial $\bR$-bundle  $\pi^{\cS'}: \cS' \times \bR \to \cS'$ for some open subset  $\cS' \subset \cS$. Moreover, if   $\cS$ is the total space of an $A'$-bundle $\pi: \cS \to N$,   with a $1$-dimensional structure group $A' = \bR$ or $S^1$ as well,  the trivial bundle $\cS' \times \bR$ is in turn  locally diffeomorphic to a Cartesian product $\pi' \circ \pi^{\cS'}: \cV \times \bR \times \bR$  over an open subset $\cV \subset N$. Because of this, from a purely local point of view, there is essentially no loss of generality in restricting to  the RT geometries on bundles over Sasaki manifolds for which the  structure groups $A$ and $A'$ are both isomorphic to $\bR$.   This observation, together with the celebrated examples provided by the $4$-dimensional  Kerr metrics,    is the main motivation for focusing on   the  class of Lorentzian manifolds  equipped with the following structure. 
 \par
  \begin{definition} \label{Kerrmanifoldsdef}A {\it Kerr  structure of  dimension $n = 2(k+1)$} is a quadruple 
 $$\cM = (M = \cS\times \bR , \pi: \cS \to N, (J, g_o),   \cH)\ ,$$ 
 given by:
 \begin{itemize}[leftmargin = 22pt]
 \item[(a)]  An   $n$-dimensional  manifold $M$,  identifiable with  the Cartesian product   $M = \cS \times \bR$, where $\cS$ is a regular Sasaki manifold    that is  an $\bR$-bundle $\pi: \cS \to N$  over a quantisable K\"ahler manifold $N$ as described  in (b);  
 \item[(b)] A quantisable K\"ahler manifold $(N, J, g_o)$  of dimension $2 k$ with K\"ahler form  $ \o_o \= g_o(J\cdot, \cdot)$; 
 \item[(c)] A connection $ \cH \subset T\cS$ on the $\bR$-bundle $ \pi^\cS: \cS \to N$, which is the kernel distribution of    a   connection $1$-form $\theta^{\cS}$ satisfying
 $d \theta^{\cS} =  \pi^* \o_o $. 
 \end{itemize}
 A manifold $M$ as in (a), equipped with the above-defined  Kerr structure $\cM$,  is   called a {\it Kerr manifold}.   By a slight abuse of language,  we also refer to any    connected open subset   $M' = \cS \times I$, with   $I \subset \bR$,   of a Kerr manifold $M = \cS \times \bR$ as a ``Kerr manifold''.
  \end{definition}
Being a (trivial) bundle over a regular Sasaki manifold,   any  Kerr manifold  $M = \cS \times \bR$ is naturally endowed with its standard  RT geometry $\cQ_o$ as defined   in \S \ref{optquant}. For brevity,  the compatible   metrics of $\cQ_o$ are   referred to as {\it compatible metrics of the Kerr manifold}.  
 \par
     \smallskip
      \subsection{K\"ahler potentials and    complex structures on  Kerr manifolds} \label{complexstr} Let  $M = \cS \times \bR$ be a  Kerr manifold with Kerr structure $\cM = (M = \cS\times \bR , \pi: \cS \to N, (J, g_o),   \cH)$ and underlying quantisable K\"ahler  manifold  $(N, J, g_o)$.  Let also $\qs_o$ and $\ps_o$ be the vector fields of $M$, whose flows give the right actions of  the structure groups of the bundles  $\pi: \cS \to N$ and $\pi^\cS: M = \cS \times \bR \to \cS$, respectively (here, we trivially  identify the generator $\qs_o$ of the right action on $\cS$  with a vector field on the cartesian product $M = \cS \times \bR$). Then, for  each $x = (U, r)  \in M = \cS \times \bR$, we may consider the standard identification  
 \beq \label{ident} T_{(U, r)} M \simeq T_U \cS + T_r \bR =  T_U \cS + \bR \ps_o|_r \ . \eeq
Such a  Kerr manifold  is naturally equipped with the  family of  (locally defined) complex structures,  one per each  K\"ahler potential for  $g_o$, which we  define as follows.\par
\medskip
  Consider  a local trivialisation $\cS|_{\cV} \simeq \cV \times \bR$ for the $\bR$-bundle $\pi :\cS \to N$ over an open subset $\cV \subset N$ and the corresponding trivialisation 
  $\cS|_{\cV} \times \bR \simeq \cV \times \bR^2$ of 
 $M = \cS \times \bR$. Assume also that  $\cV$ is sufficiently small so that: 
 \begin{itemize}
 \item[(a)] there exists a K\"ahler potential $\f: \cV \to \bR$ for $g_o$ and
 \item[(b)] there is a system 
  of  complex coordinate $(z^1 = x^1+ \I y^1, \ldots, z^{n-1} = x^{n-1} +\I y^{n-1})$ on the open subset $\cV$ of  the complex manifold $(N, J)$. 
  \end{itemize}
Let us also adopt the shorthand notation 
  $\f_{x^i} \= \frac{\p \f}{\p x^i}$ and  $\f_{y^i} \= \frac{\p \f}{\p y^i}$.
    One can directly verify   (see e.g. \cite[Lemma 2.7]{GGSS}) that  the 
  distribution $\cD$ of the CR structure of $\cS|_{\cV}$ (which is also identifiable with a codimension two distribution of $M$ through the identification \eqref{ident}) is spanned by the vector fields:
\beq \label{theEhat} 
\begin{split}
& \wh{\frac{\p}{\p x^i}} =  \frac{\p}{\p x^i} -  d^c \f\left(\frac{\p}{\p x^i}\right) \qs_o =  \frac{\p}{\p x^i} +  \f_{y^i}  \frac{\p}{\p u}\ , \\
 & \wh{\frac{\p}{\p y^i}} =  \frac{\p}{\p y^i} -  d^c \f\left(\frac{\p}{\p y^i}\right) \qs_o =  \frac{\p}{\p y^i} -  \f_{x^i}  \frac{\p}{\p u}
 \end{split}  \eeq
 and  the complex structure $J$ of the CR structure $(\cD, J)$ is given by the family of {\col anti-}involutive linear isomorphisms of the spaces $\cD_{(U, r)}$  defined by 
 \beq \label{theJ} J\left(\wh{\frac{\p}{\p x^i}} \right) = \wh{\frac{\p}{\p y^i}} \ ,\qquad J\left(\wh{\frac{\p}{\p y^i}} \right) = - \left(\wh{\frac{\p}{\p x^i}} \right)\ .\eeq 
 Consider now the diffeomorphism 
 $$\Phi^\f: \cV \times \bR^2 \longrightarrow \cV \times \bR^2 \ ,\qquad \F^\f(x^i,y^j, u, r) = (x^i, y^j, u, r + \f(x^i, y^j) )\ ,$$
 whose associated pushing-forward map $\F^\f_*$ is such that 
 \begin{multline*} \F^\f_*\left(\frac{\p}{\p x^i}\right) = \frac{\p}{\p x^i} + \f_{x^i} \frac{\p}{\p r}\ ,
 \qquad \F^\f_*\left(\frac{\p}{\p y^j}\right) = \frac{\p}{\p y^j} + \f_{y^j} \frac{\p}{\p r}\ ,\\   \F^\f_*\left(\frac{\p}{\p u} \right) =  \frac{\p}{\p u} \ ,\qquad 
 \F^\f_*\left(\frac{\p}{\p r} \right) =  \frac{\p}{\p r}\ .
 \end{multline*}
 Thus,  $\F^\f_*$ sends the CR structure  $(\cD, J)$   into  the CR structure
 $(\cD', J'|_{\cD'})$ of $\cV \times \bR^2$, given by the distribution $\cD'$ spanned by the  vector fields 
 $$\F^\f_*\left(\wh{ \frac{\p}{\p x^i}} \right) =   \frac{\p}{\p x^i} +  \f_{y^i}  \frac{\p}{\p u} +   \f_{x^i}  \frac{\p}{\p r}  \ ,\qquad 
 \F^\f_*\left( \wh{\frac{\p}{\p x^i} }\right) =   \frac{\p}{\p y^i}  -  \f_{x^i}  \frac{\p}{\p u} +   \f_{y^i}  \frac{\p}{\p r} \ ,$$
 and the restriction   to $\cD'$ of the  (integrable) complex structure $J'$ on $\cV \times \bR \times \bR$,  defined by 
 $$J'\left(\frac{\p}{\p x^i}\right) = \frac{\p}{\p y^i}\ ,\qquad J'\left(\frac{\p}{\p y^i}\right) = -  \frac{\p}{\p x^i}\ ,\qquad J'\left(\frac{\p}{\p u}\right) = \frac{\p}{\p r}\ ,\qquad J'\left(\frac{\p}{\p r}\right) =   - \frac{\p}{\p u}\ .$$
The pulled-back  (integrable) complex structure $J^{(M, \f)} \= F^{\f *}(J')$ on  $\cS|_{\cV} \times \bR \simeq \cV \times \bR^2$ is called
the  {\it  complex structure  canonically determined by the K\"ahler potential $\f$}. \par
\medskip
By construction, the CR structures $(\cD, J)$ of the  hypersurfaces $\cS|_{\cV} \times \{r_o\}  \subset  \cS|_{\cV} \times \bR$, $r_o \in \bR$,   coincide with the CR structures which are induced 
on these hypersurfaces by   $J^{(M, \f)}$. Notice also  that the  $\bC$-valued coordinates on $\cS|_{\cV} \times \bR$ defined by 
\beq \label{ahah} (z^i \= x ^i+ \I y^i, w \= u + \I (r + \f(x,y))\eeq
are  given  by the expression of the  map $\F^\f$ in terms of the 
 (holomorphic) complex coordinates $(z^i ,w)$ for the complex manifold $\big(\cS|_\cV \times \bR, J'\big)$.
 Thus   \eqref{ahah}   are  (holomorphic) complex coordinates  for the complex manifold  $\big(\cS|_\cV \times \bR, J^{(M, \f)}\big)$. \par
\smallskip

\subsection{Null fluid gravitational fields  and quasi-Einstein metrics} \label{sect33}
Let $(M, g)$ be an $n$-dimensional Lorentzian manifold. From now on, {\it we adopt the notation $\sign(g)$ to indicate the integer $+1$ or $-1$,  
depending on whether  the signature of $g$  is mostly plus or mostly minus, respectively}. \par
\smallskip
Let  $g$ represent  a  gravitational field determined by   
a so-called {\it perfect null fluid}. From a mathematical point of view, this condition   means that the metric  $g$ is a solution of an equation of the form
\beq\label{quasi}  \Ric - \frac{1}{2} \Scal\, g = -   \L' g + {\mathsf k}\ T
\eeq
where $\mathsf{k}$  is the physical constant  ${\mathsf k} =  \frac{8 \pi G}{c^4}$ and where:
\begin{itemize}
\item $\Ric$ and $\Scal$ are the Ricci and scalar curvatures of $g$, respectively,
\item  $ \L' \= \sign(g) \Lambda$, with  $\L$ denoting  the cosmological constant of a background metric on $M$, 
\item $T$ is a symmetric tensor field of type $(0, 2)$, which represents the   energy momentum tensor
of a perfect null fluid. 
\end{itemize}
We recall that the latter condition on   $T$ means that  the tensor $ {\mathsf k} T$ has  the form 
$${\mathsf k} T = \bff\, \q \vee \q$$
for some smooth real function   $\bff: M \to \bR$   and   a $1$-form $\q$ that is null  with respect to  $g$. \par
\smallskip
Note that: 
\begin{itemize}[leftmargin = 15pt]
\item[--] The scalar curvature $\Scal$ of a metric satisfying an equation of the form  \eqref{quasi} is  necessarily constant  -- actually,  taking the $g$-trace on both sides of \eqref{quasi}, one obtains that  $\Scal = \frac{2n}{ n-2} \L'$,  so that   \eqref{quasi} is equivalent to  
\beq\label{quasi-bis}  \Ric  =  \frac{2}{n-2} \L' g + {\mathsf k} T\ .\eeq 
\item[--] The solutions to \eqref{quasi} (or  to \eqref{quasi-bis}) are analogs  of the  ``Lorentzian quasi-Einstein metrics"  defined in   \cite{SKH}, that is,  solutions to  equations of the form \eqref{quasi} for  some  {\it time-like}  $\q$.
\end{itemize}
\par
\medskip
If $M$ is a Kerr manifold  with  Kerr structure  $\cM = (M = \cS\times \bR , \pi: \cS \to N, (J, g_o),   \cH)$,  and if $\q = \pi^{\cS*} \theta$   is the pulled-back $1$-form on $M$ corresponding to the contact   form of $\cS$, we know that $\q$  is null with respect to {\it any} of the  compatible metrics   of $M$.  
 Thus, a  natural ansatz   for finding  solutions to \eqref{quasi}   is to assume that  the  unknown  $g$ is  a compatible metric   of a  Kerr manifold.  This  is  the main  motivation for considering  the following 
\begin{definition} A compatible metric $g$ of a Kerr manifold $M = \cS \times \bR$ is called {\it quasi-Einstein} if  it satisfies an equation of the form
\beq \label{qE-eq}
\Ric =   \frac{2}{n-2} \L'  g + \bff\, \q \vee \q\ ,\qquad\text{with}\  \q = \pi^{\cS*} \theta\ ,\eeq
for some  smooth real function $\bff$ and some real constant $ \L'$.
\end{definition}
We conclude this section with the following useful  remark: {\it if   ${\mathsf F}: M \to M$  is an automorphism of  $\q$ (i.e.,   a diffeomorphism such that   
$ {\mathsf F}^* \q = \q$), then
 a  metric $g$  is a solution to \eqref{qE-eq} if and only if  $g' \= {\mathsf F}^* g$  is a solution of that equation too.} \par
 \medskip
   Very simple examples of automorphisms of the $1$-form $\q$ of a Kerr manifold $M = \cS \times \bR$
are those whose  coordinate expressions have  the form
 \beq \label{bending} {\mathsf F}(x, y, u, r) = \bigg(x, y, u, r'(x,y,u,r)\bigg)\eeq
 where we  denote by  $(x,y,u)$ a set of local coordinates for $\cS$  and by $r$ the  standard  coordinate for the fiber $\bR$. 
  The (local) diffeomorphisms   of the form \eqref{bending} will be called   {\it bending diffeomorphisms}.  It is clear  that the quadruple
 \beq \label{bended} \cQ^{\mathsf F}_o = \bigg(  {\mathsf F}_*(\cW_o),  [{\mathsf F}_*(h_o)]_\pm,  {\mathsf F}_*(\cK_o), \{{\mathsf F}_* g\}\bigg)\eeq
 is still an  optical geometry on $M$ and  the   compatible metrics of $\cQ_o$  are  isometric to the compatible  metrics of  $\cQ^{\mathsf F}_o$. On the other hand, the  \sf congruence of the optical geometry $\cQ_o$ is 
 in general different from the one of $\cQ^{\mathsf F}_o$. Indeed, the first is given  by the integral curves of 
 the generator  $\ps_o = \frac{\p}{\p r}$  of the null distribution $\cK_o$, while the \sf congruence of $\cQ^{\mathsf F}_o$ is given by the integral curves of $\ps \= {\mathsf F}_*(\ps_o)$. 
 The corresponding  optical geometry  $\cQ^{\mathsf F}_o$  (resp. metric compatible with $\cQ^{\mathsf F}_o$)  is said to be {\it ${\mathsf F}$-bended}. 
  \par
 \medskip

\section{Quasi-Einstein metrics  on    $4$-dimensional Kerr manifolds}
\label{HLN result}
In \cite{HLN} the authors  determined a complete parametrisation of the local  coordinate expressions of the $4$-dimensional Lorentzian  metrics that satisfy the following  conditions: 
\begin{itemize}[leftmargin = 18pt]
\item[(a)] they are   compatible with a prescribed twisting optical  geometry, projecting onto a strongly pseudoconvex $3$-dimensional {\it embeddable} CR manifold (i.e. CR equivalent to a real  hypersurface of $\bC^2$); 
\item[(b)] they satisfy  null fluid  equations of the form  \eqref{quasi-bis}.  
\end{itemize}
The  parametrisation   in \cite{HLN} is given  in terms of   the  solutions of a  particular system of  complex p.d.e.'s.
 Building on this result,    we  give here a detailed   description of the  quasi-Einstein Lorentzian metrics,  which are compatible with the  optical geometry of a Kerr $4$-manifold. This is possible because  any such optical geometry is twisting and projecting onto  a Sasaki CR structure. Indeed, since the Sasaki CR structures are well known to be  embeddable (for instance, see \cite{BRT, Jac, EKS1}), the optical geometries we  consider in this paper  nicely  fit within the class that  is considered  in  \cite{HLN}. \par
\medskip
Throughout the following,   $M = \cS \times \bR$ is a Kerr $4$-manifold,  equipped with a Kerr structure $\cM = (M = \cS\times \bR , \pi: \cS \to N, (J, g_o),   \cH)$ and  the  corresponding  standard RT geometry $\cQ_o$. We 
 use the notation $\pi^\cS: M = \cS \times \bR  \to \cS$ and $\wc \pi = \pi \circ  \pi^\cS: M \to N$ to indicate the two  projections of $M$ onto $\cS$ and $N$, respectively.  We also recall that, given a Lorentzian metric $g$,  we define  $\sign(g)$  to be equal to   $+1$ (resp.$-1$)  if  the signature of $g$ is $(+,+,+,-)$  (resp. $(-,-,-,+)$).
\begin{theo}\label{nuw1} Let $g$ be a $\cQ_o$-compatible Lorentzian metric on  $M = \cS \times \bR$, i.e. of the form
\begin{multline}\label{comp1*}  g =\s \wc \pi^* g_o + \q \vee \bigg(\s \a \ps^*_o + \s \gd +   \frac{\s \b}{2} \q\bigg),\\ 
\text{where}\  \q \= \pi^* \theta \text{\ is the pulled-back $1$-form of the contact form $\theta$ of $\cS$}\\
 \text{(thus, with   $d\q = \wc \pi^*(\o_o)\ ,\ \o_o \= g_o(J\cdot, \cdot)$)}\end{multline}
   and assume that it has    constant scalar curvature  $\Scal =  4 \L' $ for some $\L' \in \bR$ and  has a datum  $\big(\s, \a, \gd, \b \big)$ satisfying the conditions  
 \beq \label{conditions}  \qs_o(\s) =\qs_o(\a) = \qs_o(\b) = 0\ ,\qquad  \cL_{\qs_o} \gd = 0\ .\eeq
  Let  also $\cV \subset N$ be an open subset of $N$, on which there exists a  potential $\f$ for the  K\"ahler form $\o_o =  d d^c \f$ and  for which there are  coordinates   $\xi = (x, y, u, r)$   on   $\wc \pi^{-1}(\cV) $  such that
    \begin{itemize}[leftmargin = 18pt]
      \item[--] $x$ and $y$  are  the real and imaginary parts of a holomorphic   coordinate   $z = x + \I y$ on  $\cV $, 
    \item[--]$u$ is a fiber coordinate for  the $\bR$-bundle $\cS|_{\cV} \simeq \cV \times \bR$,  so that $\qs_o = \frac{\p}{\p u}$, 
    \item[--]  $r$ is  the standard coordinate  of the factor $\bR$ of $M = \cS \times \bR$, so that $\ps_o = \frac{\p}{\p r}$.
    \end{itemize}
    This metric satisfies an equation of the form $\Ric =  \L'  g + \bff\, \q \vee \q$  for some smooth function $\bff$  on $\wc \pi^{-1}(\cV)$  if and only if,  on such a set, it   has the form
\begin{multline}\label{111}
  g =\frac{1}{4} \left(\k+\frac{r'{}^2}{ \k}\right) \wc \pi^* g_o
+ \vartheta \vee \bigg\{d r' + \underset{= - d^c \k}{\underbrace{\k_y dx - \k_x dy }}+ \\
+ \bigg(\k  + \frac{r'{}^2}{\k}\bigg) \left( t_2   dx +  t_1 dy  \right)
+ 2  \big( t_2\k       - r' t_1      \big)dx     +2 \big(t_1\k     + r' t_2  \big)dy +   \frac{1}{\D \f} \left(\k +\frac{r'{}^2}{\k}\right)\cH\, \vartheta \bigg\}\  ,
\end{multline}
where:
\begin{itemize}[leftmargin=18pt, itemsep = 8pt]
\item[(i)]$t_i =  t_i(x, y) $,   $ i = 1,2$,  are  the real and imaginary parts of a  solution $t = t_1 + \I t_2$ to 
\beq  \label{337} \p_z t -  (\p_z (\log \D \f) + t)t = 0\ ;\eeq 
\item[(ii)]   $\k = \k(x, y)$ is a  function  of the form $\k \= \sign(g) \wt \k^2$ for a    solution $\wt \k = \wt \k(x, y)$ of
  \begin{equation}\label{con2*}
\D \wt \k   =  \frac{ 2^8 \Re(m)}{(\D \f)^2} \frac{1}{\wt \k^3} + 
 \left(\frac{1}4 \D \left(\log \D \f\right )  + 3 \left(\p_z \bar{t} + \p_{\bar z} t + |t|^2\right) \right) \wt  \k +    \left( \frac{\Lambda \D \f}{6} \right) \wt \k^3  \ ,
\end{equation}
with $\L = \sign(g) \L'$ and   $m(x, y)  = m_1(x, y) + \I m_2(x, y)$   solution   to
\beq \label{cccon}  \partial_z  m - 3\bigg( \p_z \left (\log \D \f \right ) + t\bigg)m = 0\ \ \  \text{(here,  $t = t_1 + \I t_2$ is  the  function  in (i)) \ .}\eeq
\item[(iii)] denoting $f \= \log \big(\frac{\D \f}{4}\big)$, the function  $\cH$ depends only on $x, y, {\col r}$ and  is equal to the sum $\cH \= h_o + h_m  + h_t$,     in which we use the notation 
\begin{align}
&\nonumber h_o\=\frac{1}{6 (\k^{2} + r'^{2})} \bigg(
10 \Lambda \k^{3}\D \f
+ 2\Lambda \k \D \f r'^{2} + \\
& \hskip 5 cm 
+ 3\k^{2}\D (\log(\D\f))  
+ 3 \left (\k_{x}^{2}+  \k_{y}^{2} \right )
- 3 \k \D \k
\bigg)\ ,\\
\nonumber &h_t\=\frac{1}{\k^2+r'^2}\Bigg(-t_{2x} r' \k
-t_1 \k_y r'
+t_1 f_x \k^2
-t_1 f_x r'^2
-t_2 f_y \k^2
+t_2 f_y r'^2
-t_{1y} r' \k\\
\nonumber & \hspace{2.5cm}+2 \k t_2 \k_y
-2 \k t_1 \k_x
-t_2 \k_x r'
-\frac{\k^{\frac{5}{2}} \sqrt{2}\, {\mathrm e}^{\frac{f}{2}} f_x t_1}{2}
-\frac{\k^{\frac{3}{2}} \sqrt{2}\, {\mathrm e}^{\frac{f}{2}} \k_x t_1}{2 }-\\
\nonumber & \hspace{2.5cm}-\frac{\sqrt{\k}\, \sqrt{2}\, {\mathrm e}^{\frac{f}{2}} r'^{2} f_y t_2}{2 }
+\k^{\frac{3}{2}} \sqrt{2}\, {\mathrm e}^{\frac{f}{2}} r' f_y t_1
+\k^{\frac{3}{2}} \sqrt{2}\, {\mathrm e}^{\frac{f}{2}} r' f_x t_2
-\frac{\sqrt{2}\, {\mathrm e}^{\frac{f}{2}} r'^{2} \k_y t_2}{2 \sqrt{\k}}\\
\nonumber & \hspace{2.5cm}+\frac{\sqrt{2}\, {\mathrm e}^{\frac{f}{2}} r'^{2} \k_x t_1}{2 \sqrt{\k}}
+\sqrt{\k}\, \sqrt{2}\, {\mathrm e}^{\frac{f}{2}} r' \k_y t_1
+\sqrt{\k}\, \sqrt{2}\, {\mathrm e}^{\frac{f}{2}} r' \k_x t_2
+11 t_1^{2} \k^2
+t_1^{2} r'^2\end{align}
\begin{align}
\nonumber & \hspace{2.5cm}+5 t_{1x} \k^2
-5 t_{2y} \k^2
+11 t_2^{2} \k^2
+t_2^{2} r'^2
+\frac{\k^{\frac{5}{2}} \sqrt{2}\, {\mathrm e}^{\frac{f}{2}} f_y t_2}{2 }
+\frac{\k^{\frac{3}{2}} \sqrt{2}\, {\mathrm e}^{\frac{f}{2}} \k_y t_2}{2 }\\
& \hspace{2.5cm}-2 f_y t_1 r' \k
-2 f_x t_2 r' \k
+\frac{t_2 \k_y r'^{2}}{\k }
-\frac{t_1 \k_x r'^{2}}{\k }
+\frac{\sqrt{\k}\, \sqrt{2}\, {\mathrm e}^{\frac{f}{2}} r'^{2} f_x t_1}{2}\Bigg)\ , \\
&h_m\=\frac{2^6 \k  \left(\k m_1 - m_2 r' \right)}{(\D \f)^2\left(\k^{2}+r'^{2}\right)^{2}}\ , \hskip 8 cm
\end{align}
where     $t = t_1 + \I t_2$, $\k$  and $m = m_1 + \I m_2$  are  the   functions defined   in  (i) and  (ii);
\item[(iv)] $r' = r'(x, y,r)$ is a real  function having the form 
\beq\label{the radius}  r'(x, y, r) =  \k(x, y)  \tan\left(\frac{{\col\frac{1}{2}  \int_0^r  \a(x,y,s) ds}   + \f(x, y) + s(x,y)}{2}\right)\ , \eeq where $s {=} s(x, y)$  is a  freely specifiable function of $x$ and $y$, and $\k$ is  the function  in  (ii).
\end{itemize}
\end{theo}

\begin{rem}\label{Remarkone} As noted in the Introduction,  all  these metrics have algebraically special curvature tensors at all points. Moreover, from the results in \cite{HLN}, for any such metric
  the Weyl scalar  $\Psi_2 = \Psi_2(x, y, r)$ with respect to the  adapted tetrads   considered in that paper,   is given  in terms of the functions $\f$, $\k$, $r'$ and $m = m_1 + \I m_2$  by 
\beq
\Psi_2 =\frac{8\sign(g)}{  \k^2  e^{3f}} \frac{ m }{(\k - \I r')^3} \ .
\eeq
It follows that,  if  $m = m_1 + \I m_2$  is a   non-zero solution to \eqref{cccon} (resp. if $m \equiv 0$), then the corresponding metric has a   Weyl scalar $\Psi_2$ that is not identically zero   (resp. has the Weyl scalar  $\Psi_2 $ identically zero) and  the  Petrov type of its curvature is  II or D  on  the set  $\{\Psi_2 \neq 0\}$ (resp.  III, N or 0 at all points).
\end{rem}

\begin{pf}
Let  $(x,  y, u,  r)$ be   coordinates  on $\wc \pi^{-1}(\cV) $  as  in the statement and denote by ${\mathsf F}: \cS|_{\cV} \times \bR \to  \cS|_{\cV} \times \bR$ the bending  diffeomorphism  
 \beq \label{bending-1} {\mathsf F}(x, y, u, r) = \bigg(x, y, u,  \wc r = {\col\frac{1}{2}  \int_0^r  \a(x,y,s) ds} \bigg)\ .\eeq
  Under this diffeomorphism,   $\cQ_o$  is mapped  to  the optical structure   \eqref{bended} and, in particular,   the $\cQ_o$-compatible metric $g$ is mapped to the $\cQ_o^{\mathsf F}$-compatible metric $g^{\mathsf F}\= {\mathsf F}_* g$.  Since 
 \begin{multline*}{\mathsf F}_* ( \wc \pi^* g_o) = \wc \pi^* g_o\ ,\quad {\mathsf F}_* (\q) = \q \ ,\\ {\mathsf F}_* (\ps^*_o) = ({\mathsf F}^{-1})^* dr  ={\col  \frac{2}{ \a}  d \wc r - \bigg( \frac{1}{\a} \int_0^{r(x,y, \wc r)} \frac{\p \a}{\p x}\bigg|_{(x,y,s)} ds\bigg) dx  -  \bigg(   \frac{1}{\a} \int_0^{r(x,y, \wc r)}  \frac{\p \a}{\p y}\bigg|_{(x,y,s)} ds\bigg) dy} \ ,\end{multline*}
the metric  $g^{\mathsf F}$  has the form 
\begin{multline}\label{comp1***}  g^{\mathsf F} =\s \wc \pi^* g_o + \q \vee \bigg(2  \s \, d \wc r  + \s \gd' +   \frac{\s \b}{2} \q\bigg)\ , \\
 \text{where}\ \gd' \= \gd -{\col  \bigg( \frac{1}{\a} \int_0^{r(x,y, \wc r)} \frac{\p \a}{\p x}\bigg|_{(x,y,s)} ds\bigg) dx  -  \bigg(   \frac{1}{\a} \int_0^{r(x,y, \wc r)}  \frac{\p \a}{\p y}\bigg|_{(x,y,s)} ds\bigg) dy}\ .\end{multline}
By the remarks in \S \ref{sect33}, 
 $g $ is quasi-Einstein if and only if  $g^{\mathsf F} $  is. Moreover (1) $g^{\mathsf F}$  is  $\cQ^{\mathsf F}_o$-compatible and (2) $\cQ_o^{\mathsf F}$ is the standard optical geometry for the Kerr structure on $\wc \pi^{-1}(\cV)$ which is given if one  identifies  $\wc \pi^{-1}(\cV) (= \cS|_{\cV}\times \bR)$ with the cartesian product  $\cS|_{\cV}\times \bR$ via the  diffeomorphism ${\mathsf F}: \cS|_{\cV} \times \bR \to  \cS|_{\cV} \times \bR$ (instead of just using the identity map!). 
  Hence,  by replacing the original Kerr structure with the one just described, expression   \eqref{comp1***} shows that there is no loss of generality  if,  from now on,  we assume  that $g$  has  a datum $(\s, \a, \gd, \b)$ satisfying the  following more restrictive  conditions: 
  \beq \label{conditions-bis} \a \equiv 2  \ , \qquad  \cL_{\qs_o} \gd = 0\ ,\qquad \qs_o(\s) =\qs_o(\b) = 0\ .\eeq 
In fact, in order to  recover  general  results  from those  under the restricted assumption \eqref{conditions-bis},   it is  sufficient  to make the  very simple replacements 
\begin{multline*} r  \dashrightarrow 
 {\col\frac{1}{2}  \int_0^r  \a(x,y,s) ds}\qquad \text{and}\\
 \gd \dashrightarrow  \gd -{\col  \bigg( \frac{1}{\a} \int_0^{r} \frac{\p \a}{\p x}\bigg|_{(x,y,s)} ds\bigg) dx  -  \bigg(   \frac{1}{\a} \int_0^{r}  \frac{\p \a}{\p y}\bigg|_{(x,y,s)} ds\bigg) dy}\ .
 \end{multline*} \par
  \medskip
Consider now   the complex coordinates
$(z \= x + \I y, w \= u + \I \wt r)$ with  $ \wt r \= r + \f(x,y)$,  
    defined in 
\eqref{ahah}. We observe that:
\begin{itemize}[leftmargin = 15pt]
\item[--]  $(z,w)$ are holomorphic coordinates  for the {\it integrable} complex structure $J^{(M, \f)}$;  
\item[--]  $J^{(M, \f)}$ induces the CR structure $(\cD, J)$ on each hypersurface
$$\cS^{(\cs)} \= \cS|_{\cV}  \times\{\cs\} \subset  \wc \pi^{-1}(\cV) \simeq \cS|_{\cV} \times \bR\ ,\qquad {\cs \in \bR}\ ;$$
\item[--]  Each   $\cS^{(\cs)} $  is  equal to  the set of  solutions to the equation
$$u - F(z, \bar z) - \cs = 0 \qquad \text{where we use the notation}\qquad F(z, \bar z) \= \f\left(\frac{z + \bar z}{2}, \frac{z - \bar z}{2\I}\right)\ ;$$
\item[--]The tensor fields   $\wc \pi^* g_o$, $\q = \pi^{\cS*} \theta$ and  $\ps_o^* = d r$  have the form 
\begin{multline*} \hskip 1 cm \wc \pi^* g_o =\D \f (dx\vee dx  + d y\vee dy) = 4  F_{z \bar z} d z \vee d \bar z\ ,\\
 \q = du - \f_y dx + \f_x dy = du - i F_z d z + i F_{\bar z} d \bar z  \ , \qquad \ps_o^* = d  \wt r -  d F\ ;\end{multline*} 
  \item[--] The   conditions   \eqref{conditions-bis}   imply that $\s$, $\b$,  $\gd$ are  independent on  $u = \Re(w)$; 
  \item[--] The signature of $g$ is mostly plus (resp. mostly minus) if and only if $\s > 0$ (resp. $\s < 0$),  meaning that   $\sign(g) = \sign(\s)$.  
\end{itemize}
We now introduce the real  functions $\cP$, $\cH$ and the complex function $\cW$ defined by 
\beq \label{theP} \cP\= \sqrt{ 2 |\s| F_{z \bar z}} \ ,\qquad \cH \= \frac{ \b F_{z\bar z}}2\ ,\qquad  \Re( \cW d z) \=\frac \gd {4}  - \frac{ d F}{2 }\ . \eeq
Recalling that $dr = d \wt r - d\f = d \wt r - dF$, a direct check shows that, after a multiplication by $\sign(g)$,   any  metric  \eqref{comp1*} satisfying \eqref{conditions-bis} (in particular,  with $\a \equiv  2 $) has the form 
   \begin{equation}\label{shearfreemetric}
   \sign(g) g =  2{\cP}^2\left(d z \vee  d \bar z  + \frac{1}{2 F_{z \bar z}}\q\vee \bigg(d \wt r  + 2 \Re\big( \cW d z\big) +  \frac{\cH}{2 F_{z \bar z}} \q\bigg)\right)\ , 
\end{equation}
namely   the  form of the metrics considered in \cite{HLN, GSS} (\footnote{For a correct comparison, keep in mind   that the $\lambda$ of   the formulas in   \cite{HLN, GSS} is actually  equal to $\lambda = \frac{1}{2 F_{z \bar z}}\q$.}). 
In other words,  the mostly plus metrics $\sign(g) g $ satisfying  \eqref{conditions-bis}  are nothing but   the    metrics  considered  by Hill, Lewandowski and Nurowski  satisfying  the additional assumption that  the   three
 functions  \eqref{theP}  are  independent on  $u$.  Combining this condition   with  the results     in   \cite{HLN}, in  \cite{GSS} it is proved that $ \sign(g) g$  is a solution to an equation   of the form  \eqref{quasi-bis}  with cosmological  constant $ \Lambda = \sign(g) \Lambda'$  if and only if the  functions \eqref{theP} are as follows: 
 \begin{align}
\label{11}  &\cP(x, y, r)  = \frac{p(x,y)}{\cos(\frac{r + F(x, y) + s(x,y)}{2})} \ , \\
 \label{eqx-y} & \cW(x, y, r) = \I \left( - t-f_z + \frac{2 p_z}{p} \right) e^{-\I ( r+ F(z, \bar z) + s(z, \bar z))}  - \I t +  \I \left(- t -f_z +\frac{  2 p_z}{p}\right)  + s_z\ ,\\
\nonumber &
\cH(x, y, r)  =2\Re\Bigg( \frac{m}{p^4}\e^{2\I\left(r + F(z, \bar z) +s\right )}+
\Bigg( 
\frac{3m + \bar{m}}{p^4} + \frac{2}{3}  \Lambda p^2 
+ \frac{p_z\, p_{\bar z}  - pp_{z\bar z}}{p^2} 
- 2t p_{\bar z} - \bar{t} \frac{p_z}{p} \\
\nonumber & \hskip 6.2 cm 
+ \frac{3}{2}  t_{\bar z} +f_{\bar z} t 
+ \frac{1}{2} f_z \bar{t} + \frac{5}{2} t \bar{t} 
+ \bar{t}_z + f_{z\bar{z}} \Bigg)\mathrm{e}^{\I( r+ F(z, \bar z) + s)}+\\
\nonumber & \hskip 5 cm 
+ \frac{3m + 3\bar{m}}{p^4} + 2 \Lambda p^2 
+ \frac{2p_z\, p_{\bar z}  - 2pp_{z\bar z}}{p^2} 
- 3( t\frac{p_{\bar z}}{p}+\bar{t} \frac{p_z}{p} )+ \\
 \label{33} & \hskip 6 cm 
+ \frac{5}{2}(\bar{t}_z + t_{\bar z}) + 6t \bar{t}
+\frac{3}{2}(f_z \bar{t} + f_{\bar z} t) + 2  f_{z\bar{z}}
\Bigg)\ ,
\end{align}
where:
\begin{itemize}[leftmargin = 15pt]
\item  $s = s(z, \bar z)$  is a freely specifiable smooth real valued functions of $z$ and $\bar z$; 
\item    $p = p(z, \bar z) \neq 0$ is a nowhere vanishing  real function  (\footnote{The constraint    $p(z, \bar z)\neq 0$ at all points follows from the fact that,  by   \eqref{11}, if  $p(x_o, y_o) = 0$ at some $(x_o, y_o)$,  then $ \cP$  would be zero  on a non empty set  $\{(x_o, y_o)\times \bR  \times (a, b) \}$  and    \eqref{shearfreemetric} would    not be  a metric. })
and $t = t_1(z, \bar z) + \I t_2(z, \bar z)$ and  $m = m_1(z, \bar z) + \I m_2(z, \bar z)$ are   complex function,  both  depending just on $z$ and $\bar z $,   constrained 
by the following  system of p.d.e.'s: 
\begin{align}
\label{c1} & \p_z t -  (\p_z f + t)t = 0\ ,\\
\label{c2} &  \partial_z  m - 3(\p_z f  + t)m = 0\ ,\\
\label{c3} &  \bigg(2\p_z  \p_{\bar z}   - f_{\bar z} \p_z - f_z \p_{\bar z} + \frac{1}{2}|f_z|^2 - \frac{3}{2} f_{z \bar z}    - \frac{3}{2}\left(\p_z \bar{t} + \p_{\bar z} t + |t|^2\right)\bigg)p = \frac{m + \bar{m}}{p^3} + \frac{2}{3} \Lambda p^3,
\end{align}
where  $f \= \log\left (F_{z\bar{z}} \right )$.
\end{itemize}
For a given   $s = s(z, \bar z)$ and a triple $(t, m, p)$    solving  \eqref{c1} -- \eqref{c2},  we set 
 \beq \label{bigchange}  \k(z, \bar z)  \=  \sign(g) \frac{ 8 p^2(z, \bar z)}{\D F(z, \bar z)}\ ,\quad  r ' (z, \bar z, r) \=  \k(z, \bar z)   \tan\left(\frac{ r + F(z, \bar z) + s(z,\bar z)}{2}\right)\ .\eeq
After a  (long and tedious) sequence of  straightforward manipulations  (for the benefit of the reader,    a sketch of  these manipulations  is given   in  Appendix \ref{appendix}), one gets   that  a necessary and sufficient condition for  $\sign(g) g$  to have  the form  \eqref{shearfreemetric}
 with     $\cP$, $\cW$ and $\cH$  as  in \eqref{11} -- \eqref{33} (thus, a strictly plus metric satisfying a quasi-Einstein equation)  is that    $g$   has the form  \eqref{111} with $r'$ as in  \eqref{the radius}  with  $\a \equiv 2$.  As we have pointed out before, the  expression for a compatible  metric with   $\a$  arbitrary function independent on $u$ is obtained starting from  the one we just described via the  overall replacement $r \dashrightarrow {\col\frac{1}{2}  \int_0^r  \a(x,y,s) ds}$. This leads to   \eqref{111}  in full generality.\par
\par
\smallskip
Since \eqref{c1} and \eqref{c2} are  the conditions \eqref{337} and \eqref{cccon}, in order to conclude, we just need to show that  the constraint   \eqref{c3} on the triple  $(t, m,p)$ is equivalent to  the  constraint \eqref{con2*} on the triple $(t, m, \k)$ where, being $\sign(g) \k > 0$,  the function $\k$ can be  written  as  $ \k = \sign(g) \wt \k^2$ for some smooth function $\wt \k$. 
In order to check this, let  us   write \eqref{c3} as 
\beq \label{con}
 2\frac{  p_{z\bar z}}{p}   - f_{\bar z} \frac{p_z}{p} - f_z \frac{p_{\bar z}}{p}	  + \frac{1}{2}|f_z|^2 - \frac{3}{2} f_{z \bar z}   - \frac{3}{2}\left(\p_z \bar{t} + \p_{\bar z} t + |t|^2\right) =  \frac{m + \bar{m}}{p^4} + \frac{2}{3} \Lambda p^2.
\eeq
We  now observe that, from $p^2 =    \frac{ \sign(g) \k F_{z \bar z} }{2} =  \frac{ \sign(g) \k  e^f}{2} $,   
\begin{multline*}
2\frac{  p_{z\bar z}}{p}=\p_{z\bar{z}}\left (\log p^2\right )+\frac{1}{2} \p_z \left(\log p^2 \right )\p_{\bar{z}} \left(\log p^2 \right )=\\
= \p_{z\bar {z}}\left(\log( \sign(g) \k) \right )
+f_{z\bar {z}}+\frac{1}{2}\left ( \frac{\k_z}{\k} +f_z\right )\left(\frac{\k_{\bar{z}}}{\k} +f_{\bar{z}} \right )=\\
=\frac{\k_{z\bar{z}}}{\k}-\frac{1}{2}\frac{\k_z}{\k}\frac{\k_{\bar{z}}}{\k}+f_{z\bar {z}} 
+\frac{1}{2} \frac{\k_z}{\k}f_{\bar{z}}
+\frac{1}{2}f_z\frac{\k_{\bar{z}}}{\k}  
+\frac{1}{2}  |f_z|^2\ .
\end{multline*}
Plugging this in  \eqref{con} and using the relation  $\L = \sign(g) \L'$, we get 
\beq \label{con1}
 \frac{\k_{z\bar{z}}}{\k}-\frac{1}{2}\frac{\k_z}{\k}\frac{\k_{\bar{z}}}{\k}
 - \frac{1}{2} f_{z \bar z}   - \frac{3}{2}\left(\p_z \bar{t} + \p_{\bar z} t + |t|^2\right) = \frac{4 (m + \bar{m}) e^{-2f}}{\k^2 } + \frac{1}{3} \sign(g) \Lambda' \k e^f \ .
\eeq
 Since   $\sign(g) \k=   \wt \k^2$,   $\k_z=2 \sign(g) \wt \k \wt \k_z$,  $ \k_{z\bar{z}}=\sign(g)(2  \wt \k_{\bar{z}}\wt \k_z+2  \wt \k \wt \k_{z\bar{z}})$, 
it  follows that  \eqref{con1}  is equivalent to 
 \beq \label{con2}
2\frac{\wt \k_{z\bar{z}}}{\wt \k} 
 - \frac{1}{2} f_{z \bar z}   - \frac{3}{2}\left(\p_z \bar{t} + \p_{\bar z} t + |t|^2\right) = \frac{4 (m + \bar{m}) e^{-2f}}{\wt \k^4 } + \frac{1}{3} \Lambda \wt \k^2 e^f\ . 
\eeq
Recalling that  $f = \log(F_{z \bar z}) = \log\left(\frac{\D\f}{4}\right)$, this is equivalent to   \eqref{con2*}, as claimed.   
 \end{pf}
 \begin{rem} The above arguments show that \eqref{con2*} is actually equivalent to    \eqref{con1}, i.e. to the  following  elliptic equation on  $\k = \sign(g) \wt \k^2$: 
 \beq \label{con111}
  \frac{\D \k}{ \k}-\frac{1}{2}\frac{\k_x^2 + \k_y^2}{\k^2}
 - \frac{1}{2} \D(\log \D \f)   - 6\left(\p_z \bar{t} + \p_{\bar z} t + |t|^2\right) - \frac{2^9 \Re (m) }{\k^2( \D \f)^2}- \frac{1}{3}  \Lambda' \k  \D \f= 0 \ .
\eeq
 \end{rem}
 \par
\medskip
 \section{A new  large class of examples of   null fluid gravitational fields}
By the results in \S \ref{HLN result},  the $\cQ_o$-compatible  Lorentzian metrics on a Kerr manifold  $M$  corresponding to   null fluid gravitational fields  are locally   of the form  \eqref{111}. 
In this section  we  focus on  the following subclass  of   gravitational fields, which we will shortly see  naturally include and generalise the well known Ricci flat Kerr metrics.
\begin{definition} We say that a Lorentzian metric on a Kerr manifold,  as in Theorem  \ref{nuw1},   {\it belongs to  the enlarged Kerr family } if the corresponding functions $\k$, $t$ and $m$ satisfy the following conditions for some constant $\bfB \neq 0$:
\beq  \label{kerrfamily} \k = \bfB \f\ ,\qquad t = t_1 + \I t_2 \equiv 0\ ,\qquad m = m_1 + \I m_2\ \text{is nowhere vanishing}\ .
\eeq
\end{definition}
The third condition in  \eqref{kerrfamily}  and  \eqref{cccon} imply that,  locally, $m$ is a solution to the equation  $ \partial_z  \log \big(\frac{ m_1 + \I m_2}{(\D \f)^3} \big)= 0$
 for some holomorphic  logarithm. If  $\cV \subset N$ is a simply connected  domain where such a holomorphic logarithm exists, and  if  $(x_o, y_o) \in \cV$ is a fixed point, then  all real   functions $m_1 = \Re(m)$ and $m_2= \Im(m)$  can be described as follows: 
 \beq \label{form}  m_1  = (\D\f)^3(\bfm + \bfA) \ ,\qquad m_2 = (\D\f)^3( \bfa + \bfA^*)\ ,\eeq
 where: 
 \begin{itemize}[leftmargin = 15pt]
 \item[--] $\bfa$ and $\bfm$ are two freely specifiable real constants; 
 \item[--] $\bfA(x, y)$ is an arbitrary harmonic function on $\cV$ which is  $0$ at $(x_o, y_o)$; 
  \item[--] $\bfA^*(x, y)$ is the unique  harmonic function,   which is  $0$ at $(x_o, y_o)$ and is such that  the complex function  $\bfA(x, y) + \I \bfA^*(x, y)$ is anti-holomorphic (\footnote{If  we denote by $\wt \bfA$   the unique harmonic conjugate of $\bfA$ vanishing at $(x_o, y_o)$, then 
{\col  $\bfA^*(x, y) := \wt\bfA(-y, x)$}.
  }). 
 \end{itemize}
The next proposition and the following remarks motivate our interest for this class of metrics. In  what follows  $\cV \subset N$ is a simply connected domain on which  $ m_1$ and  $m_2$ are  as in \eqref{form} and    $\bfB$ is  the constant which determines   $\k$  as in  \eqref{kerrfamily}. We also assume that  the local identification between the Kerr manifold  $M|_{\cV}$ and  the cartesian product $\cS|_{\cV} \times \bR$ is determined by an  appropriate bending diffeomorphism (not necessarily the identity map),  so that  the  function    \eqref{the radius} and the $1$-form $dr'$ are just $r'(x, y, r) = r$ and    $dr'  = dr$.  
  \begin{prop} \label{theprop} A metric with scalar curvature $\Scal = 4 \L'$ is in  the enlarged Kerr family  on $\wc \pi^{-1}(\cV) \simeq \cS|_{\cV} \times \bR \subset M$  with associated  constant $\bfB \neq 0$  if and only if there are
  \begin{itemize}
  \item[(a)]  two real constants  $\bfa$,  $\bfm$,  
  \item[(b)] 
  a harmonic function $\bfA(x, y)$  vanishing  at $(x_o, y_o)$  {\col and such that the complex function  $(\D\f)^3\bigg((\bfm + \bfA) + \I ( \bfa + \bfA^*)\bigg)$ is nowhere vanishing}, 
   \item[(c)] a solution  $\f = \f(x, y)$  to the equation 
\beq \label{con111**}
 \f  \D \f - \frac{1}{2}(\f_x^2 + \f_y^2)
 - \frac{\f^2}{2} \D(\log \D \f)  -  \D\f\left(\frac{2^9 (\bfm + \bfA)  }{\bfB^2} + \frac{1}{3}  \Lambda' \bfB \f^3\right)= 0 \ ,
\eeq
\end{itemize}
such that $ g =   \bfB \wc g$ with  
 \begin{equation}\label{111****}
  \wc g =    \frac{1}{4 } \left( \f^2+\frac{r^2}{ \bfB^2}\right) \frac{\D \f}{\f}(dx^2 + dy^2) + 
 \vartheta \vee \bigg( d\left(\frac{r}{\bfB} +  u\right)  + \big( - \frac{1}{2}  + \wt \b_o\bigg)\, \vartheta \bigg)   \  ,
\end{equation}
where  we use the notation
\begin{multline} 
\wt \b_o \=  2^{6}  \bigg(\frac{-7(\frac{\bfm}{\bfB^2} + \frac{\bfA}{\bfB^2}) \f^2 -  \f( \frac{\bfa}{\bfB^2} + \frac{\bfA^*}{\bfB^2}) \frac{r}{\bfB} - 8(\frac{\bfm}{\bfB^2} + \frac{\bfA}{\bfB^2})(\frac{r}{\bfB})^2}{\f( \f^2+\left(\frac{r}{\bfB}\right)^2)}\bigg) +\\
+\Lambda' \bfB\bigg( \frac{(5\sign(g)  -  {\col 1})}{3}  \f^2
+  \frac{\sign(g)}{3}\left( \frac{r}{\bfB} \right)^2 \bigg) 
 \ .
\end{multline}
 \end{prop} 
 \begin{pf} Assuming that $g$ has the form \eqref{111} and that  \eqref{kerrfamily} is satisfied, from the identity  $\q = d u + d^c \f $ and \eqref{form}  we may write $g$ as 
\begin{multline}\label{111*}
  g =  \frac{1 }{4 } \left( \bfB \f+\frac{r^2}{ \bfB\f}\right) \wc \pi^* g_o
+ \vartheta \vee \bigg\{d r  -\bfB d^c \f +  \frac{1}{\D \f}   \left(\bfB \f + \frac{r^2}{\bfB \f}\right)(h_o + h_m)\, \vartheta \bigg\} = \\
=  \frac{1 }{4 } \frac{\bfB^2  \f^2 +r^2}{ \bfB} \frac{\D \f}{\f} (dx^2 + dy^2)  + 
 \vartheta \vee \bigg\{d\left( r + \bfB u\right)  + \bfB\bigg( - 1 + \frac{\bfB^2 \f^2 + r^2}{\bfB^2 \f \D \f}(h_o + h_m)\bigg)\, \vartheta \bigg\}  \  ,
\end{multline}
with  
\begin{align}
&\nonumber h_o\=\frac{1}{\bfB^2 \f^{2} + r^{2}} \bigg(
\frac{5}{3} \Lambda\bfB^3  \f^{3}\D \f
+ \frac{1}{3} \Lambda \bfB  \f \D \f r^{2} + \\
& \hskip 5 cm 
+ \frac{1}{2}\bfB^2 \f^{2}\D (\log(\D\f))  
+ \frac{1}{2}  {\bfB^2} \left (\f_{x}^{2}+  \f_{y}^{2} \right )
- \frac{1}{2}  \bfB^2 \f \D \f
\bigg)\ ,\\
&h_m\=  \frac{ \bfB \f \D \f }{\bfB^2 \f^2+r^{2}}  \frac{2^6 \left(\bfB \f  (\bfm + \bfA) -  ( \bfa + \bfA^*) r \right)}{\bfB^2 \f^2+r^{2}}\ .
\end{align}
 Since $t = 0$,  by \eqref{con111}  the K\"ahler potential $\f = \frac{1}\bfB \k$ is constrained  by   
\beq \label{con111*****}
 \bfB^2  \f \D \f  - \frac{1}{2} \bfB^2 (\f_x^2 + \f_y^2) 
  - 2^{9}  \D\f (\bfm + \bfA)-  \frac{1}{3}    \Lambda' \bfB^3 \f^3  \D \f= \frac{1}{2}  \bfB^2  \f^2 \D(\log \D \f)  \ . 
\eeq
This equality implies that  $h_o$  is also given by  
\beq 
 h_o\=\frac{\bfB \f \D \f }{\bfB^2 \f^{2} + r^{2}} \bigg(
\frac{(5\sign(g)  -  {\col 1} )}{3}\Lambda'\bfB^2  \f^2
+ \frac{\sign(g)}{3} \Lambda' r^{2} 
+ \frac{1}{2} \bfB
    - 2^{9}  \frac{\bfm + \bfA}{\bfB \f} 
\bigg)\ ,
\eeq
so  that   
\begin{multline}\label{cris}   - 1 +\frac{\bfB^2 \f^2 + r^2}{\bfB^2 \f \D \f}(h_o + h_m)  {=}\\
{=} -1   - 2^{9}  \frac{\bfm + \bfA}{\bfB^2 \f} 
 + \frac{2^6 \left(\bfB \f  (\bfm + \bfA) -  ( \bfa + \bfA^*) r \right)}{\bfB(\bfB^2 \f^2+r^{2})} +\Lambda'\bigg(\frac{(5\sign(g)     -  {\col 1} )}{3}\bfB  \f^2
+ \frac{\sign(g)}{3\bfB} r^{2}  \bigg) +\frac{1}{2}{=} \\
{=} -\frac{1}{2}   - 2^{9}  \frac{\bfm + \bfA}{\bfB^2 \f} 
 + \frac{2^6 \left(\bfB \f  (\bfm + \bfA) -  ( \bfa + \bfA^*) r \right)}{\bfB(\bfB^2 \f^2+r^{2})} +\Lambda'\bigg(\frac{(5\sign(g)     -  {\col 1} )}{3}\bfB  \f^2
+ \frac{\sign(g)}{3\bfB} r^{2}  \bigg) {=} \\
{=}   -\frac{1}{2}   +2^{6}  \bigg(\frac{-7(\bfm + \bfA)\bfB^2 \f^2 -  \bfB \f( \bfa + \bfA^*) r - 8(\bfm + \bfA)r^{2}}{\bfB^2 \f(\bfB^2 \f^2+r^{2})}\bigg) +\\
+\Lambda'\bigg(\frac{(5\sign(g)     -  {\col 1} )}{3}\bfB  \f^2
+ \frac{\sign(g)}{3\bfB} r^{2}  \bigg) 
 \ .
\end{multline}
This implies the claim. 
\end{pf}
Proposition \ref{theprop} shows that the quasi-Einstein metrics with a  scalar curvature $\Scal = 4 \L'$ belonging to  the enlarged Kerr family are locally parametrised by: 
\begin{itemize}
\item[(a)] The freely specifiable constants $\bfB \neq 0$, $\bfa$, and $\bfm$, 
\item[(b)] The freely specifiable harmonic function $\bfA$ vanishing at a fixed  point {\col (subjected just  to the very mild constraint that   $\big((\bfm + \bfA) + \I ( \bfa + \bfA^*)\big)$ is not identically vanishing)}, 
\item[(c)]  The solutions  to  \eqref{con111*****}.
\end{itemize}
Note also that the metric $\wc g$ defined in \eqref{111****} is isometric to a metric $\wt g$  with $\Scal = 4 \bfB \L'$ and associated with the {\col the new set of} constants $\bfB' = 1$, $\bfa' = \frac{\bfa}{\bfB^2}$, $\bfm' = \frac{\bfm}{\bfB^2}$ and with the function $\bfA' =\frac{\bfA}{\bfB^2}$ (in order to pass from $\wc g$ to $\wt g$ it suffices to  make the replacement $r  \dashrightarrow \bfB r$). Hence {\it up to an  homothety and a change of signature, for any quasi-Einstein metric of the enlarged Kerr family   the corresponding constant $\bfB$ can be assumed to be  $\bfB = 1$}. 
\par
\smallskip
We claim that the set of solutions of (c) (and, consequently,   the corresponding set of metrics) is  very large. In fact, in case   $\bfm = 0$ and  $\bfA \equiv 0$ (so that   $\bfA^* \equiv 0$ as well),  \eqref{con111**} reduces to 
\beq \label{con111***}
  \frac{\D \f}{\f} - \frac{1}{2 \f^2}(\f_x^2 + \f_y^2)
 - \frac{1}{2} \D(\log \D \f) + \frac{\f\D \f}{3}  \Lambda' \bfB = 0 \ .
\eeq
If we limit ourselves to the case   $\L' = 0$ and we introduce the function $\psi = \frac{\D \f}{|\f|}$,  the solutions $\f$ to \eqref{con111***} appear to be  in bijection with the pairs $(\f, \psi)$  that satisfy 
\beq\label{1313} \D \f - \sign(\f) \psi \f= 0\ ,\qquad  \D(\log \psi) - \sign(\f)  \psi = 0\ .\eeq
As it is observed in the proof of \cite[Thm. 4.1]{GGSS},  we may always select coordinates $(x, y)$ on $\cV$,  in which a solution $\psi$ to the second equation takes  the form $ \frac{8}{( 1- \sign(\f)(x^2 + y^2))^2}$.  This yields that in such coordinates  $\f$ is  either a solution to 
\beq \label{dych}
\begin{split} & \D \f + \frac{8 \f}{( 1+ x^2 + y^2)^2} = 0\qquad \text{in case} \qquad \f < 0\ \qquad \text{or to }\\
&  \D \f - \frac{8 \f}{( 1- x^2 - y^2)^2} = 0\qquad \text{in case}\qquad \f > 0\ .
\end{split}
\eeq
Explicit expressions for a countable set of solutions to these  conditions and corresponding explicit   expressions for     metrics with  $\L' = 0$,    $\bfB = 1$,  $\bfm = 0$ and  $\bfA \equiv 0$ have been given in \cite[\S\ 6]{GGSS}. There it is shown that  the  metrics of this kind with $\f$ solution to  \eqref{dych}    (which is equivalent to \eqref{con111***}) are Ricci flat  and constitute a class that  properly include  all metrics of  Kerr black holes.  As we observed  above, replacing $\bfB = 1$ by any other non-zero value $\bfB' \neq 1$ determines a homothetic  metric  with a possibly opposite signature, thus Ricci flat as well. \par
{\col  We now claim that a basic application  of the Implicit Function Theorem in Banach spaces yields  the  existence of  a multitude of  local solutions to     \eqref{con111**}, with  non zero constant   $\bfm$ and/or non zero  scalar curvature  $4 \L'$,  and  with non-identically vanishing functions $\bfA\neq 0 $.  We also claim that such solutions determine an infinite family of pairwise non isometric metrics that are solutions to  the Einstein equation with  non-vanishing null fluid energy momentum tensors.  Details for all this  are given in  the next two subsections.\par
\smallskip
\subsection{On the existence of  solutions with non zero  $\bfm, \L', \bfA$}
  From now on,  we assume   the non-restrictive hypothesis that the constant $\bfB$ is $\bfB = 1$ and  we denote by  
   $(\bfm_o, \L'_o) $ a pair of real constants different from the pair $(0, 0)$ and by  $\bfA_o$   a non-identically vanishing harmonic function  with $\bfA_o(0,0) = 0$ and  with the property that,  for at least one constant $\bfa_o$,    the  function   $(\bfm + \bfA) + \I ( \bfa + \bfA^*)$ with   $(\bfa, \bfm, \bfA) = (\bfa_o, \bfm_o, \bfA_o)$  is nowhere vanishing.  We also denote by  $\bD = \{x^2 + y^2 < 1-\ve\} $, $0< \ve < 1$,   a fixed disk in $\bR^2$ centred at the origin and of radius  $1 - \ve$ strictly smaller than $1$.  To simplify the forthcoming expressions,  we also introduce the notation
   $$a_o \= 2^9 (\bfm_o + \bfA_o) \ ,\qquad b_o \= \frac{1}{3}  \Lambda_o'\ .$$
  We observe that under the assumptions 
    $$\bfB = 1\ ,\qquad \bfm = t\bfm_o\ ,\qquad \bfA = t\bfA_o\ ,\qquad\L' = t \L'_o\qquad \ \text{for some}\ t \in \bR\ ,$$ 
     the   equation  \eqref{con111**}   
      takes the form 
      \beq \label{con111**-ter}
 \f  \D \f - \frac{1}{2}(\f_x^2 + \f_y^2)
 - \frac{\f^2}{2} \D(\log \D \f)  -  t \D\f\left(a_o  + b_o  \f^3\right)= 0 \ ,
\eeq
which clearly reduces to  the equation  \eqref{con111***} with $\L' = 0$ if we assume that  $t = 0$.   Considering only  functions  $\f$  of constant  sign  over the closed disk $\overline \bD$ and by a  straightforward manipulation of \eqref{con111**-ter}, one can directly check  that the  solutions  to \eqref{con111**-ter}
 are in one-to-one correspondence with the maps $(v, \f): \overline \bD \to \bR^2$   satisfying   (here, $\pm 1 = \sign \f$)
\beq \label{514}
\begin{split}
&\D\f  \mp  e^v \f = 0\ ,\\
& \f  \D v \pm e^v \left(- \f + 2 (a_o  + b_o \f^3)t\right) = 0\ .
 \end{split}
 \eeq
Now,  for fixed   $k \geq 2$ and   $\a \in (0,1)$, let   $\gB_1$,  $\gB_2$,  $X$ be  the Banach spaces  
\beq 
\begin{split}
 \gB_1 &\= \cC^{k, \a}(\overline{\bD}) \times \cC^{k, \a}(\overline{\bD}) \ ,\\
\gB_2 &\= \cC^{k-2, \a}(\overline{\bD}) \times  \cC^{k, \a}(\p \bD) \times \cC^{k-2, \a}(\overline{\bD)}   \times \cC^{k, \a}(\p \bD) \ ,\\
X &\= \cC^{k, \a}(\p \bD) \times \cC^{k, \a}(\p \bD) \times \bR\ .
\end{split}
\eeq
 Finally, let $G^{(+)}$ and $G^{(-)}$  be the operators defined by 
\begin{multline}  G^{(\pm)}: \gB_1 \times X \to \gB_2\ ,\qquad
G^{(\pm)}\big((v, \f),( {\bf v}, {\bf f}, t)\big) \=\\
=  \bigg(\f \D v \pm e^v(- \f + 2 (a + b \f^3) t), (v - {\bf v})|_{\p \bD}, \D \f \mp e^v \f,(\f - {\bf f})|_{\p \bD}\bigg)\ .
\end{multline}
The $\cC^{k,\a}$-solutions $(v,\f): \overline \bD \to \bR$ to \eqref{514} with  $\sign(\f)$ identically equal to  $+ 1$ or $-1$ (they correspond to  $\cC^{k,\a}$ solutions of constant sign to \eqref{con111**-ter}) 
 are   the elements  in  $\gB_1 $  with 
 \beq \label{518} G^{(\pm)}\bigg((v, \f), \gs\bigg) = (0, 0, 0, 0)\qquad \text{where}\ \gs \=  \big({\bf v} = v|_{\p \bD}, {\bf f} =  \f|_{\p \bD}, t\big)\ . \eeq
 An infinite family of  solutions  to   \eqref{518} with $t = 0$ (corresponding to solutions to  \eqref{con111***} with $\L' = 0$)
  can be  found as follows by the  explicit solutions $\f_o: \bD \to \bR $  determined  in \cite[\S 6]{GGSS}. We recall that,  denoting by $\f_o^{(+)}$ (resp.  $\f_o^{(-)}$)  a solution given in  that paper with  $\sign(\f^{(+)}_o) = + 1$ (resp.  $\sign(\f^{(-)}_o) = - 1$),     any such a map  $\f^{(\pm)}_o$ is a  solution to  \eqref{dych} and is therefore also a solution to the first equation  of the system  \eqref{514} for $t = 0$, provided that the function $v$ is  equal to   $ v_o^{(\pm)} \=  \log\left( \frac{8}{( 1\mp (x^2 + y^2))^2}\right)$. 
  Note also that, at the same time, $v = v^{(\pm)}_o$ is solution to the second equation of \eqref{514} for $t = 0$ regardless on what is  the considered function  $\f$.  This means that,  considering   one of the two functions $v_o^{(\pm)} :\bD \to \bR$ as above,  a corresponding solution  $\f^{(\pm)}_o:\bD \to \bR$  to \eqref{dych} and the pair and triple 
  $$u _o \= (v^{(\pm)}_o, \f^{(\pm)}_o)\ ,\qquad \gs_o\= \big({\bf v}_o \= v^{(\pm)}_o|_{\p \bD}, {\bf f}_o \=  \f^{(\pm)}_o|_{\p \bD}, t = 0\big)$$
 we have that  $(u_o, \gs_o)$ is  an element of $\gB_1 \times X$ for which  \eqref{518} holds.
   \par
  We now observe that $G^{(\pm)}: \gB_1 \times X \to \gB_2$ is continuously Frechet differentiable at any such $(u_o, \gs_o) \in \gB_1 \times X$. Furthermore,  we claim  that,  if   
 we use the notation $$D_{(u_o, \gs_o)}G^{(\pm)}: \gB_1 \times X \to \gB_2$$ for  the  Frechet derivative of $G^{(\pm)}$ at $(u_o, \gs_o)$, the  associated restricted linear operator
 $$D_{(u_o, \gs_o)}G^{(\pm)}\big|_{\gB_1}: \gB_1 \longrightarrow \gB_2$$
 is  invertible.  Indeed, computing explicitly the Frechet derivative  at $(u_o, \gs_o)$, one can directly check that the invertibility of such linear operator  is equivalent to  the existence  and uniqueness of  $\cC^{k, \a}$ solutions $(\wh v, \wh \f)$ on $\overline \bD$ to the        Dirichlet problem  
 \beq \label{palla}
 \begin{split}
 & \D \wh v \mp e^{v^{(\pm)}_o} \wh v = h_1\ ,\qquad \text{with}\quad   \wh v|_{\p \bD} = {\bf h_1}\ ,\\
 & \D \wh \f \mp e^{v^{(\pm)}_o} \wh \f \mp \f^{(\pm)}_o    e^{v^{(\pm)}_o} \wh v = h_2\ ,\qquad \text{with}\quad   \wh \f|_{\p \bD} = {\bf h_2}
 \end{split}
 \eeq
 for any  choice of $(h_1,  {\bf h}_1, h_2, {\bf h_2}) \in \gB_2$.  In the case of  $G^{(+)}$,  such desired existence and uniqueness   follows from the fact that, for  both  equations in \eqref{palla}, the  coefficient of the $0$-th order term $ - e^{v^{(+)}_o}$ is non-positive, a property  that yields the  conclusion  by  a well known    result  on  elliptic operators.  In  the case of   $G^{(-)}$, this argument is no longer applicable. However,   expressing  the unknown  function $\wh v$ as a product of the form  $\wh v(x, y) = f_o(x, y) \wt v(x, y)$ with $f_o(x, y) \= - \frac{1- x^2 - y^2}{1 + x^2 + y^2}$, the solutions to the first differential problem  in  \eqref{palla} are in  bijections  with the solutions  $\wt v$ to  
  \beq \label{palla-bis}
  \D \wt v + \frac{2}{f_o} \left(\frac{\p\wt v}{\p x} \frac{\p f_{o}}{\p x} +  \frac{\p \wt v}{\p y} \frac{\p f_{o}}{\p y}\right) = \frac{h_1}{f_o}\ ,\qquad \text{with}\quad   \wt v|_{\p \bD} = \frac{\bf h_1}{f_o|_{\p \bD}}\ . 
 \eeq
 For this new elliptic problem,  the desired existence and uniqueness follows  immediately from the identically vanishing of the  $0$-th order term (see also  \cite[Lemma 6.1]{GGSS}). An almost identical  argument  yields the desired existence and uniqueness for the second  equation in  \eqref{palla} as well. 
 \par
 \smallskip
 Due to the invertibility of 
 $D_{(u_o, \gs_o)}G^{(\pm)}\big|_{\gB_1}$,   the Implicit Function Theorem (see e.g. \cite[Thm.17.6]{GT}) applies and  yields the existence of solutions to   
 \eqref{518} for any triple $\big({\bf v} = v|_{\p \bD}, {\bf f} =  \f|_{\p \bD}, t\big) \in X$ which is sufficiently close to $\big({\bf v}_o \= v^{(\pm_o)}|_{\p \bD}, {\bf f}_o \=  \f^{(\pm_o)}|_{\p \bD}, t = 0\big)$.  It follows that,   for any triple of the form $\big({\bf v}_o \= v^{(\pm_o)}|_{\p \bD}, {\bf f}_o \=  \f^{(\pm_o)}|_{\p \bD}, t \big)$,  there exists a  $\cC^{k, \a}$ solution on $\bD$ to  \eqref{con111**-ter} for any $t$ sufficiently close to $0$. Actually, using  classical results on regularity of solutions of elliptic equations, one can directly check that any such  solution is of class $\cC^\infty$, and therefore determines a   Lorentzian metric in  the  enlarged Kerr family   with $(\bfm , \L')  \neq (0,0)$,  $\bfA \not \equiv 0$  which is arbitrarily  close to  the explicit  Ricci flat metric   of the class considered in \cite[\S 6]{GGSS}, associated with the above  chosen function  $\f_o: \bD \to \bR $. } \par
 \medskip
 \subsection{Metrics of the enlarged Kerr family with   non vanishing  source terms}
  {\col The  expression   for a metric of the enlarged Kerr family  $g = {\bfB} \wc g$ with $\wc g$ as in \eqref{111****}  shows that {\it locally}  any such metric $g$ is compatible not only for  the  RT geometry  $\cQ_o$ of $M \times \cS$, but also for a second optical geometry, which we now describe.  As usual, in what follows and with no loss of generality, we  restrict to the cases with  $\bfB =1$, since all other metrics  are  homothetic to  these. \par
  \smallskip
  Consider a metric $g = \wc g$ as in \eqref{111****},  in  an  open set  $\cU = \cV \times \bR^2 \subset M$ where are defined the coordinates $(x,y, u, r)$, with $(x, y)$ coordinates for $\cV \subset N$ and such that we can write   $\qs_o = \frac{\p}{\p u}$ and $\ps_o = \frac{\p}{\p r}$. On $\cU$ we may  consider the frame field $(\wc E_1, \wc E_2, \wc\ps_o, \wc\qs_o)$ 
  defined by   \beq \label{adap}\wc E_1 \= \frac{\p}{\p x} + \f_x\left( \frac{\p}{\p u} - \frac{\p}{\p r}\right), \wc E_2 \= \frac{\p}{\p y} - \f_x\left( \frac{\p}{\p u} - \frac{\p}{\p r}\right),\ \wc\ps_o \= \ps_o = \frac{\p}{\p r},\ \wc\qs_o \= \left( \frac{\p}{\p u} - \frac{\p}{\p r}\right)\ .\eeq
Its dual coframe field $(\wc E_1, \wc E_2, \wc\ps_o^*, \wc\qs_o)$   is given by 
 $$\wc E^1 = dx, \qquad \wc E^2 = dy, \qquad \wc\ps_o^* = d(r + u),\qquad  \wc\qs_o^* = \q = du + d^c \f\ .$$
  Using these frame and coframe fields,  the metric   \eqref{111****}   can be written as 
 \beq \label{buona}
 \begin{split}
& g  =  \wc\s  \wc g_o(E_i,  E_j)  \wc E^i \vee \wc E^j +\wc\qs^*_o \vee \left(\wc\s \wc\a \wc\ps_o^* + \wc \s  \wc\g^i  \wc g_o(E_i, E_k) \wc E^k  + \frac{\wc\s \wt \b}{2} \wt \qs^*_o\right)\ \text{where} \\ 
&
\wc g_o \= \D \f (dx^2 + dy^2)\ ,\  \wc\s \=  \frac{\f^2+ r^2}{4\f } \ ,\  \wc\a \= \frac{1}{\s}\ ,
\quad \wc\g^i \equiv 0\ ,\ \wc\b  \= \frac{1}{\s}  \left(- 1 + 2\wt \b_o\right)\ .
\end{split}
\eeq
This means that $g = \wc g$  can be considered as  compatible also  with  an optical geometry $\wc \cQ_o \neq \cQ_o$,  in which $\qs_o = \frac{\p}{\p u}$ is replaced by $\wc \qs_o = \frac{\p}{\p u} - \frac{\p}{\p r} $ and for which  \eqref{adap} is an  adapted frame  (in the sense of \cite{AGSS1}). Considering  $g = \wc g$ as compatible with such new optical geometry has    the  advantage of  having identically vanishing  components $\wc \g^i$ when written in the form \eqref{buona}. Indeed,  condition  $\wc \g^i \equiv 0$  allows great simplifications in the forthcoming formulas. \par
\smallskip 
Let  $\wc \nabla$ be the Levi-Civita connection of  the metric \eqref{buona} and denote by  $\GGa A B C$ the Christoffel symbols of $\wc \nabla$ with respect to the frame field $(X_A)$ defined  in \eqref{adap} (\footnote{By ``Christoffel symbols'' we mean  the functions  $\GGa A B C$ appearing  in the expansions $\n_{X_A} X_B = \GGa A B C X_C$}).   Expressions for the Christoffel symbols  in terms of the   datum of a  metric, which is compatible with an optical structure,  have been determined in \cite{AGSS1} (see also \cite{AGSS}). These expressions can be immediately  used to  determine  the Christoffel symbols and, in turn,  the  components of the Ricci tensors of  the metrics \eqref{buona}.   In the following  we give the complete   list of such Christoffel symbols using  the  same  notational  conventions of \cite{AGSS1}. For instance,    we denote by $\GGa i j m$    the components  of  the covariant derivative  $\n_{\wc E_i} \wc E_j$  in the direction of  the vector fields $\wc E_m$, by    $\GGa i j {\wc \ps_o}$       the component of  the vector field $\n_{\wc E_i} \wc E_j$  in the direction of $\wc \ps_o$ 
  and so on. We also set  $E_1 \= \frac{\p}{\p x}$, $E_2\= \frac{\p}{\p y}$, so that  the standard complex structure of $\bC = \bR^2$  can be defined as the complex structure such that  $J(E_1) = E_2$ and $J(E_2) = -E_1$. We finally denote  $g_{ij} \= \wc g_o(E_i, E_j)$,  $\o_{ij} \= \wc g_o(J E_i,  E_j)$.   Now,  recalling that $\wc \g^i \equiv 0$ and that all  functions $\wc \s, \wc \a, \wc \b$ are independent of the coordinate $u$,  from the formulas in  \cite[Prop. 3.1]{AGSS1}   and a few   straightforward manipulations, for a metric \eqref{buona} we have 
  \begin{align}
\label{746ter}
\GGa i j m &=
  g^{mk} g_o(\n^o_{E_i}   E_j, E_k)  
    + \frac{1}{2 \s}  \wc E_i(\s) \d_j^m +  \frac{1}{2 \s}  \wh E_j(\s) \d_i^m 
 - \frac{g_{ij}}{2 \s}  g^{mk} \wc E_k( \s) \ ,\\%
 \GGa i j {{\wc \ps_o} }  & =    2 g_{ij} \wt \b_o {\wc \ps_o} (\s) \ , \qquad
 \GGa i j {{\wc \qs_o} }  =  - \frac{  \o_{ij}}{2}  -  g_{ij}  {\wc \ps_o} (\s)\ , \\[7 pt]
 \GGa i {{\wc \ps_o} } m & = \GGa  {{\wc \ps_o} } i m =  \frac{1}{4 \s}\left(J_i^m + 2  {\wc \ps_o} (\s) \d_i^m\right) \ , \ \ 
 \GGa i {{\wc \ps_o} } {{\wc \ps_o} }=  \GGa {{\wc \ps_o} } i {{\wc \ps_o} } = \GGa i {{\wc \ps_o} } {{\wc \qs_o} }=  \GGa  {{\wc \ps_o} } i {{\wc \qs_o} } = 0\ ,\\[7 pt]
 \nonumber
 \GGa i {{\wc \qs_o} } m &=  \GGa  {{\wc \qs_o} } i m  =  \frac{J_i^m } {4 \s} \left(- 1 + 2\wt \b_o\right)-\frac{1}{2 \s} {\wc \ps_o} (\s)\d_i^m\ ,\\
 %
&\GGa i {{\wc \qs_o} } {{\wc \ps_o} } =  \GGa  {{\wc \qs_o} } i  {{\wc \ps_o} }  =   \wc E_i\left(\wt \b_o\right)  \ ,
 \qquad
 \GGa i {{\wc \qs_o} } {{\wc \qs_o} } =  \GGa  {{\wc \qs_o} }  i {{\wc \qs_o} } =  0  \ ,\\[7 pt]
 %
 \GGa {{\wc \ps_o} }  {{\wc \ps_o} }  m  & =  
 \GGa {{\wc \ps_o} }   {{\wc \ps_o} } {{\wc \ps_o} }  = 
 \GGa {{\wc \ps_o} }   {{\wc \ps_o} } {{\wc \qs_o} }  =  0 \ , \\[7pt]
 \GGa {{\wc \ps_o} }  {{\wc \qs_o} }  m & =  \GGa {{\wc \qs_o} }   {{\wc \ps_o} }  m = 0 \ ,
\qquad 
 \GGa {{\wc \ps_o} }   {{\wc \qs_o} } {{\wc \ps_o} }  =  \GGa  {{\wc \qs_o} }  {{\wc \ps_o} }  {{\wc \ps_o} }  =  {\wc \ps_o} \left( \wt \b_o\right)  \ , \qquad
 \GGa {{\wc \ps_o} }   {{\wc \qs_o} } {{\wc \qs_o} }   =   \GGa  {{\wc \qs_o} }  {{\wc \ps_o} }   {{\wc \qs_o} } = 0 \ ,
 \end{align}
 \begin{align}
\label{5.2555}
\GGa {{\wc \qs_o} }  {{\wc \qs_o} }  m & = - \frac{g^{mk}}{2 \s}  \wc E_k\left( \wt \b_o\right)  \ ,\qquad
\GGa {{\wc \qs_o} }   {{\wc \qs_o} } {{\wc \ps_o} }  = 2 \left(- 1 + \wt \b_o\right){\wc \ps_o} \left(\wt \b_o\right) 
   \ ,
\qquad  \GGa {{\wc \qs_o} }   {{\wc \qs_o} } {{\wc \qs_o} }  =  -  {\wc \ps_o} \left(\wt \b_o\right) \ .
  \end{align}
  From these expressions,  we may  compute explicitly all   components of the Ricci tensor $\Ric$  in the frame field \eqref{adap}. In fact, using just definitions, one can  check  that  the    component  $\Ric({\wc \qs_o} , {\wc \qs_o} )$  of the Ricci tensor  is given in terms of  the above Christoffel symbols  by (the crossed out terms are surely vanishing because of the above list of Christoffel symbols): \begin{multline}  \label{5.2666} \Ric({{\wc \qs_o} }{{\wc \qs_o} })\  \ {=}  R_{m {\wc \qs_o}  {\wc \qs_o} }{}^m + R_{{\wc \ps_o}  {\wc \qs_o}  {\wc \qs_o} }{}^{{\wc \ps_o} } = \\
 =  \wc E_m(\GGa {{\wc \qs_o} } {{\wc \qs_o} } m) - {\wc \qs_o} (\GGa m {{\wc \qs_o} } m) - \GGa m {{\wc \qs_o} } i \GGa {{\wc \qs_o} } i m -  \xcancel{\GGa m {{\wc \qs_o} } {{\wc \ps_o} }  \GGa {{\wc \qs_o} } {{\wc \ps_o} } m} -  \xcancel{\GGa m {{\wc \qs_o} } {{\wc \qs_o} }  \GGa {{\wc \qs_o} } {{\wc \qs_o} } m}  +\\
  + \GGa {{\wc \qs_o} } {{\wc \qs_o} } i \GGa {m} i m +  \GGa {{\wc \qs_o} } {{\wc \qs_o} } {{\wc \ps_o} }  \GGa {m} {{\wc \ps_o} } m + \GGa {{\wc \qs_o} } {{\wc \qs_o} } {{\wc \qs_o} }  \GGa {m} {{\wc \qs_o} } m + \\
 +{\wc \ps_o} (\GGa {{\wc \qs_o} } {{\wc \qs_o} } {{\wc \ps_o} }) - {\wc \qs_o} (\GGa {{\wc \ps_o} } {{\wc \qs_o} } {{\wc \ps_o} }) - \xcancel{\GGa {{\wc \ps_o} } {{\wc \qs_o} } i \GGa {{\wc \qs_o} } i {{\wc \ps_o} }} -  \GGa {{\wc \ps_o} } {{\wc \qs_o} } {{\wc \ps_o} }  \GGa {{\wc \qs_o} } {{\wc \ps_o} } {{\wc \ps_o} } - \xcancel{ \GGa {{\wc \ps_o} } {{\wc \qs_o} } {{\wc \qs_o} }  \GGa {{\wc \qs_o} } {{\wc \qs_o} } {{\wc \ps_o} } } +\\
  + \xcancel{\GGa {{\wc \qs_o} } {{\wc \qs_o} } i \GGa {{\wc \ps_o} } i {{\wc \ps_o} } }+ \xcancel{ \GGa {{\wc \qs_o} } {{\wc \qs_o} } {{\wc \ps_o} }  \GGa {{\wc \ps_o} } {{\wc \ps_o} } {{\wc \ps_o} } }+ \GGa {{\wc \qs_o} } {{\wc \qs_o} } {{\wc \qs_o} }  \GGa {{\wc \ps_o} } {{\wc \qs_o} } {{\wc \ps_o} } \ . 
\end{multline}
Since  $g = \wc g$ is quasi-Einstein and satisfies the equation  
  \beq \label{qE-eq-bis}
\Ric  -  \L'  g = \bff\, \q \vee \q  \ ,\eeq
 the function $\bff$ (which is the only term needed to determine  the  source term   $ T = \bff\, \q \vee \q $) can be determined  observing  that, from  $\q(\wc \qs_o)  = \wc \qs^*_o(\wc \qs_o)  = 1$ and \eqref{qE-eq-bis}, we have  
\beq \label{source} \bff =  \Ric(\wc \qs_o, \wc \qs_o)  -  \L'  g(\wc \qs_o, \wc \qs_o)  =  \Ric(\wc \qs_o, \wc \qs_o)  -  \L'  \left( - \frac{1}{2}  + \wt \b_o\right) .\eeq 
Plugging   \eqref{5.2666} into \eqref{source} and using  \eqref{746ter} -- \eqref{5.2555}, one can get the explicit expression for the energy-momentum tensor  $ T = \bff\, \q \vee \q $ 
in terms of any choice of the data ${\bfa}$, $\bfm$, ${\bfA}$, $\L'$ and of any solution   $\f$ of  \eqref{con111**}. \par
The expressions for  $\bff$ and $T$,  which are  obtained in this way,   are  involved but, nonetheless,  fully explicit.  At a first sight, it is hard to believe that   $\bff$  is  identically vanishing for {\it all} possible choices for   ${\bfa}$, $\bfm$, ${\bfA}$, $\L'$. Actually,  it is reasonable  to  expect that,  for generic choices of such parameters,  the  corresponding metric is a solution to the Einstein equations  with  non vanishing source terms.  This very sensible  expectation can be 
rigorously established, checking that the first (or a higher order) derivative of $\bff$ with respect to one of the freely specifiable  parameter, say  $\bfm$, is not identically zero at  $\bfm_o = 0$, ${\bfA}_o = 0$, $\L'_o = 0$ and with some $\f = \f_o$  corresponding to one of the known Ricci flat metric, say  a Kerr metric. 
 We  have made this check computing the third derivative $\frac{\p^3 \bff}{\p \bfm^3} |_{ (\bfm = 0, {\bfA} = 0, \L' = 0, \f = \f_o)}$, which is one of the easiest to be determined. Just for the sake of brevity,  we omit the details of  this tedious but very straightforward computation.
}
\par
\medskip
{\col We conclude this section indicating a simple way  to determine numerically metrics of the enlarged Kerr family, corresponding to parameters }  $\bfB= 1$, $\bfa$, $\bfm$ and $\L'$ arbitrary and $\bfA \equiv 0$. Following the same approach  described above, {\col under these choices of the parameters}  the solutions to  \eqref{con111**}  are in bijection with the pairs of functions  $(\phi, \psi = e^{\wt \psi})$ solving a system of the form
  \beq \label{bis}
  \begin{split}
  & \D \f - \sign(\f) e^{\wt \psi} \phi = 0\ ,\\
& 
 \D \wt  \psi - e^{\wt \psi} \bigg(\frac{c_1}{\f}+ \sign(\f)+   c_2 {\col\f^2}\bigg)= 0\ ,
 \end{split}
\eeq
 where   the constants $c_i$, $i = 1,2$  (which are determined by $\L'$, $\bfa$ and $\bfm$),  are  possibly non-zero.  If we  limit our discussion to the solutions $(\f = \f(\gr), \psi = e^{\wt \psi(\gr)})$   depending just on  $\gr = \sqrt{x^2 + y^2}$,  the system  \eqref{bis} 
simplifies into
 \beq \label{bis*}
 \begin{split}
  & \frac{d^2 \f}{d \gr^2} + \frac{1}{\gr} \frac{d \f}{d\gr}  - \sign(\f) e^{\wt \psi} \phi = 0\ ,\\
& 
 \frac{d^2 \wt \psi}{d \gr^2} + \frac{1}{\gr} \frac{d \wt \psi}{d\gr}  - e^{\wt \psi} \bigg(\frac{c_1}{\f}+ \sign(\f)+   c_2 {\col\f^2}\bigg)= 0\ .
 \end{split}
\eeq
 By standard results on systems of ordinary differential equations,  any such   system  admits smooth solutions  on  sufficiently small neighbourhoods   $(\gr_o - \ve, \gr_o + \ve)$ of any  point $\gr_o \neq 0$. This  yields  a large class of  (local) $\cC^\infty$ solutions to \eqref{con111**}  for any choice of the cosmological $\L'$ and hence also a corresponding class of $\cC^\infty$ quasi-Einstein metrics in the enlarged Kerr family with non-zero cosmological constants. We did not succeed in finding    expressions  for such solutions in some closed   form, but  it is manifest that explicit solutions can be easily  determined numerically. {\col An    expression for the source term  $ T = \bff\, \q \vee \q $ can be determined as discussed  above and can be  determined numerically as well. These computations give  also a way to  test  numerically which -- among the several known (classical or quantum) energy conditions  -- are satisfied. Further investigations on this topic are left to future work.}
 \par
 \medskip
\section{Existence of quasi-Einstein     lifts  for   Sasaki  CR  $3$-manifolds of class $\cC^\infty$}
Let  $(\cD, J)$ be  a strongly pseudoconvex CR structure   on a $3$-dimensional manifold $\cS$. We focus on its optical lifts  over  $M = \cS \times \bR $  that satisfy the  quasi-Einstein condition, namely solutions of \eqref{qE-eq} for a fixed cosmological constant $\L = \sign(g) \L'$.  Proving (or disproving) the local existence  of  quasi-Einstein optical lifts of this kind 
has two main consequences: on the one hand, it enables the determination of CR invariants via the isometric invariants of the associated Lorentzian lifted metrics;  on the other hand, explicit or numerical methods for solving equations of the form \eqref{qE-eq} for prescribed strongly pseudoconvex CR structures provide valuable tools for studying gravitational fields generated by null fluids.
\par
\medskip
 In \cite{HLN}, Hill, Lewandowski and Nurowski obtained quite  remarkable  results  on strongly  pseudoconvex  $3$-dimensional CR manifolds  with   quasi-Einstein optical lifts.  In particular they proved the following  
 \begin{theo}[$\text{\cite[Thm. 2.5]{HLN}}$] \label{existence}  For  any real analytic strongly pseudoconvex CR $3$-di\-men\-sion\-al manifold and any choice  of a    constant $\Lambda$,   there locally exists a quasi-Einstein optical  lift (with mostly plus signature)  of Petrov type II or D with the prescribed cosmological constant $\L$.
 \end{theo} 
 The proof of this result relies crucially on the following properties of CR $3$-manifolds:
 \begin{itemize}[leftmargin = 18pt]
 \item[(a)]  Any strongly pseudoconvex, real-analytic CR $3$-manifold is locally embeddable and can be identified with a real-analytic submanifold of $\bC^2$.
 \item[(b)]  In holomorphic coordinates on $\bC^2$, the system of equations characterising the quasi-Einstein optical lifts of a CR $3$-manifold as in (a) is real-analytic and, by standard results on integrable real-analytic systems of p.d.e.'s,   locally solvable. 
 \end{itemize} 
 However, in \cite[Rem. 3.25]{HLN} the authors conjectured that the above existence result for quasi-Einstein optical lifts of type II or D  should hold under weaker regularity assumptions, provided the CR $3$-manifold is embeddable.  In support of this expectation, in \cite[Cor. 3.26]{HLN}  it is shown that, under a much weaker  hypothesis than real analyticity,   there do exist optical lifts with curvatures  of type III,  N  or 0 for embeddable CR $3$-manifolds. Furthermore,  the conjecture  was partially confirmed in \cite{GSS}, where, using results on logistic elliptic equations, the existence of a globally defined quasi-Einstein optical lift  of the desired Petrov types was established under the following hypotheses:
\begin{itemize}[leftmargin = 18pt]
\item[(1)] The underlying  CR $3$-manifold   is  $\cC^\infty$ and regular Sasaki (hence, also locally embeddable) and projects onto a  K\"ahler surface $(N, g_o, J)$ diffeomorphic to $\bR^2$;
\item[(2)] The   K\"ahler metric $g_o$ of $N$ is such that both limits $\lim_{|x| \to \infty} g_o$ and  $\lim_{|x| \to \infty} - \log(vol_{g_o})$  exist and are strictly positive;  
\item[(3)]  The prescribed cosmological constant $\Lambda$  for  the optical lift is  positive. \par\smallskip
\end{itemize}

In this concluding section, we combine Theorem \ref{nuw1} with a classical result on non-linear elliptic equations to improve the results of \cite{GSS} and confirm  Hill, Lewandowski and Nurowski's conjecture for $\cC^\infty$   $3$-dimensional Sasaki CR manifolds. Specifically, we establish the following.

\begin{cor} Let  $\L $ be an arbitrary real constant, and let $(\cS, \cD = \ker \theta, J)$ be a regular Sasaki $3$-manifold,  fibering over a quantisable K\"ahler $2$-dimensional manifold $(N, g_o, J)$. For any  sufficiently small open subset $\cV \subset N$, there exists a quasi-Einstein optical lift $g$ on $ \pi^{-1}(\cV) \times \bR \subset \cS \times \bR$ with  scalar curvature  $\Scal =  4 \Lambda'$, where $\L' = \sign(g) \L$,   and whose curvature tensors are of Petrov type II or D at all points.
\end{cor}  
\begin{pf}  By Theorem \ref{nuw1}  the class of the quasi-Einstein optical lifts  of the form \eqref{comp1*}   for the  CR structure  $(\cD, J)$ on a set of the form $\cS|_{\cV}$, $\cV \subset N $,  is non-empty if and only if the family of triples of $\cC^\infty$ functions $(t,  m, \wt \k)$, 
solving the  corresponding system   on $\cV$
\beq\label{thebigsystem}
\begin{split}
& \p_z t -  (\p_z (\log \D \f) + t)t = 0\ ,\qquad   \partial_z  m - 3\bigg( \p_z \left (\log \D \f \right ) + t\bigg)m = 0\ ,\\
& \D \wt \k   = \left( \frac{ 256 \Re(m)}{(\D \f)^2}\right) \frac{1}{\wt \k^3} + 
 \left(\frac{1}4 \D \left(\log \D \f\right )  + 3 \left(\p_z \bar{t} + \p_{\bar z} t + |t|^2\right) \right) \wt  \k +    \left( \frac{\Lambda \D \f}{6} \right) \wt \k^3
\end{split}
\eeq
is non-empty. Since the first two equations are independent on $\wt \k$ and admit local solutions on any open  subset of $\bR^2$ (the first admits the trivial solution $t \equiv 0$, and the second is linear in $m$),  the existence of triples  $(t,  m, \wt \k)$ solving \eqref{thebigsystem} reduces to the following question: given a $\cC^\infty$ K\"ahler potential $\f$ for the K\"ahler manifold $(\cV, g_o, J)$, a  cosmological constant $\L$ and two $\cC^\infty$ solutions $t$, $m$ to the first two  equations,  does there exist  a local $\cC^\infty$ solution $\wt \k$ to the third equation in   \eqref{thebigsystem}? Without loss of generality we may assume $\Re(m) \equiv 0$   and $\Im(m)$ nowhere vanishing, in which case the third equation takes the form 
$\D \wt \k = {\mathbf h}(x,y, \wt \k)$
 for a well-defined $\cC^\infty$ function ${\mathbf h}$     on $\cV \times \bR$. By a classical result on  non-linear  elliptic equations of this kind  (see e.g. \cite[p.\ 373]{CH}) and standard properties  of    elliptic equations,  a $\cC^\infty$ solution to the last equation  in \eqref{thebigsystem}  exists on any sufficiently small open subset $\cV' \subset \cV$. The conclusion is then a consequence of  Remark \ref{Remarkone} and the hypothesis that     $m$ is  nowhere vanishing.
\end{pf}
 \par
\medskip
\appendix
\section{The proof   that \eqref{shearfreemetric}  is equivalent to   \eqref{111}}
\label{appendix}
From
$ r ' \=  \k  \tan\left(\frac{ r + F + s}{2}\right)$,   we may  express  the coordinate $\wt r = r + \f = r + F$ in terms of $r'$as 
$ \wt r   = 2 \arctan\left( \frac{ r'}{\k} \right) - s$. 
Since $ F_{z \bar z} =  \frac{1}{4} \D \f$, the following equalities are immediate: 
\begin{align}
\label{A1} & g_o = \D \f (dx^2 + d y^2) = 4 F_{z \bar z} d z \vee d \bar z\ ,\\
& d\wt r  = dr + dF =  \frac{2 }{1 + \frac{r'{}^2}{\k^2}} \bigg( \frac{1}{ \k}d r' - \frac{r'}{\k^2}\left( \k_x d x +\k_y d y\right)\bigg) - s_x d x - s_y dy\ ,\\
\label{A3} &\frac{\cP^2}{F_{z \bar z}} =  \frac{ 4 \cP^2}{\D F} =  \frac{1}{2} \frac{8p^2}{\D F \cos^2\left( \frac{r + F + s}{2}\right)}=  \frac{\sign(g) \k}{2} \frac{ 1}{\cos^2\left( \frac{r +F+ s}{2}\right)} =  \frac{\sign(g)}{2}  \k \left( \frac{r'{}^2}{\k^2} + 1 \right)  \ ,\\
 \label{A4} &\frac{\cP^2}{F_{z \bar z}} d \wt r =  \sign(g) \left(d r'    -   \left(\frac{   \k_x r'}{\k} + \frac{s_xr'{}^2}{2\k} + \frac{\k s_x}{2} \right)  d x -  \left(\frac{   \k_y r'}{\k}  + \frac{s_y r'{}^2}{2\k} + \frac{\k s_y}{2} \right)  dy\right)\ .
\end{align}
Using \eqref{A1} -- \eqref{A4},  \eqref{11} and \eqref{eqx-y},  the product between \eqref{shearfreemetric} and $\sign(g)$ becomes
\begin{align}
\nonumber  g & =  \frac{1}{4} \left( \k+  \frac{r'{}^2}{ \k}\right) g_o  +   \q\vee \bigg\{   d r'    -   \left(\frac{   \k_x r'}{\k} + \frac{s_xr'{}^2}{2\k} + \frac{\k s_x}{2} \right)  d x - \\
\nonumber & \hskip 7.5 cm -  \left(\frac{   \k_y r'}{\k}  + \frac{s_y r'{}^2}{2\k} + \frac{\k s_y}{2} \right)  dy +  
\end{align}
\begin{align}
\nonumber & \hskip 0.3 cm  + \frac{1}{2} \left(\k+  \frac{r'{}^2}{\k} \right)  2  \Re\bigg(\bigg(   \I \left( - t+ {\col f_z} + \frac{2 p_z}{p} \right)\left (  e^{-\I(\wt r+ s)}+1 \right )   - \I t  + s_z \bigg) d z\bigg) + \\
\label{A5} & \hskip 9.5 cm + {\col \frac{1}{\D \f}}  \left( \k+  \frac{r'{}^2}{\k} \right)   \cH  \q\bigg\} \ ,
\end{align}
In this formula,  the term $\Re\bigg(   \I \left( - t+ {\col f_z} + \frac{2 p_z}{p} \right)\left (  e^{-\I(\wt r+ s)}+1 \right )   - \I t  + s_z \bigg) $ can be written as 
\begin{multline*}
\Re\bigg(   \I \left( - t_1-\I t_2- \frac{f_x}{2}+ \I\frac{ f_y}{2} + \frac{ p_x}{p}-\I \frac{ p_y}{p}  \right)\left (  \cos(\wt r+s)-\I\sin(\wt r+s)+1 \right ) \\
  - \I \left(t_1+\I t_2 \right )  + \frac{s_x}{2}-\I \frac{s_y}{2} \bigg) \left ( dx+\I dy\right )\bigg)=\\
   =  t_2\cos(\wt r+s)dx-\frac{ f_y}{2}\cos(\wt r+s)dx + \frac{ p_y}{p}\cos(\wt r+s)dx  + \\
      -\sin(\wt r+s)) t_1dx- \sin(\wt r+s) \frac{f_x}{2}dx +\sin(\wt r+s)) \frac{ p_x}{p}dx    + t_2dx-\frac{ f_y}{2}dx + \frac{ p_y}{p} dx 
  + t_2dx  + \frac{s_x}{2}dx  +\\
     + t_1\cos(\wt r+s) dy + \frac{f_x}{2}\cos(\wt r+s) dy   - \frac{ p_x}{p}\cos(\wt r+s) dy   + \\
       +\sin(\wt r+s) t_2 dy  -\sin(\wt r+s)\frac{ f_y}{2}dy   +\sin(\wt r+s)\frac{ p_y}{p} dy   
  + t_1 dy  + \frac{f_x}{2} dy   - \frac{ p_x}{p} dy    + t_1dy   +  \frac{s_y}{2} dy   \ .
\end{multline*}
 By classical trigonometric relations, 
$\cos(\wt r+s)=\frac{\k^2-r'^2}{\k^2+r'^2}$, $ \sin(\wt r+s)=\frac{2 r'\k}{\k^2+r'^2}$, so  that  
 the previous expression becomes 
\begin{multline*}
    t_2\frac{\k^2-r'^2}{\k^2+r'^2}dx
    -\frac{ f_y}{2}\frac{\k^2-r'^2}{\k^2+r'^2}dx 
    + \frac{ p_y}{p}\frac{\k^2-r'^2}{\k^2
    +r'^2}dx  + \\
      -\frac{2r'\k}{\k^2+r'^2} t_1dx
      - \frac{2r'\k}{\k^2+r'^2} \frac{f_x}{2}dx 
      +\frac{2r'\k}{\k^2+r'^2} \frac{ p_x}{p}dx    
      + t_2dx-\frac{ f_y}{2}dx 
      + \frac{ p_y}{p} dx 
  + t_2dx  
  + \frac{s_x}{2}dx  +\\
     + t_1\frac{\k^2-r'^2}{\k^2+r'^2} dy + \frac{f_x}{2}\frac{\k^2-r'^2}{\k^2+r'^2} dy   - \frac{ p_x}{p}\frac{\k^2-r'^2}{\k^2+r'^2} dy   + \\
       +\frac{2r'\k}{\k^2+r'^2}t_2 dy  -\frac{2r'\k}{\k^2+r'^2}\frac{ f_y}{2}dy   +\frac{2r'\k}{\k^2+r'^2}\frac{ p_y}{p} dy   
  + t_1 dy  + \frac{f_x}{2} dy   - \frac{ p_x}{p} dy     
   + t_1dy   +  \frac{s_y}{2} dy.   
\end{multline*}
Multiplying the above expression with $  \frac{\k^2+r'{}^2}{\k} $,  through some tedious but straightforward manipulation, the above expression simplifies into
\begin{multline*}
    \bigg( t_2\frac{\k^2-r'^2}{\k}
    -f_y\frac{\k^2-r'^2}{2\k} 
    +p_y\frac{\k^2-r'^2}{p\k}  
      -2r' t_1
      -f_xr' 
      +p_x\frac{2r'}{p}     \\
      +\frac{\k^2+r'{}^2}{\k} t_2
      -\frac{\k^2+r'{}^2}{\k}\frac{ f_y}{2} 
      +\frac{\k^2+r'{}^2}{\k} \frac{ p_y}{p}  
  + \frac{\k^2+r'{}^2}{\k}t_2 
  + \frac{\k^2+r'{}^2}{\k}\frac{s_x}{2}  \bigg)dx+
  \end{multline*}
  \begin{multline*}
     +\bigg( t_1\frac{\k^2-r'^2}{\k} 
     + f_x\frac{\k^2-r'^2}{2\k}   
      -p_x\frac{\k^2-r'^2}{p\k }   
       +2r' t_2  
       -f_yr'   
        +p_y\frac{2r'}{p}    
  +\frac{\k^2+r'{}^2}{\k} t_1   \\
  +\frac{\k^2+r'{}^2}{\k} \frac{f_x}{2}   
   -\frac{\k^2+r'{}^2}{\k} \frac{ p_x}{p}      
   + \frac{\k^2+r'{}^2}{\k}t_1   
   +  \frac{\k^2+r'{}^2}{\k}\frac{s_y}{2} \bigg)dy.   
\end{multline*} 
Due to this,  the following sum (appearing  in \eqref{A5}) 
\begin{multline*}   -   \left(\frac{   \k_x r'}{\k} + \frac{s_xr'{}^2}{2\k} + \frac{\k s_x}{2} \right)  d x 
-  \left(\frac{   \k_y r'}{\k}  + \frac{s_y r'{}^2}{2\k} + \frac{\k s_y}{2} \right)  dy +  \\
  + \frac{1}{2} \left(\k+  \frac{r'{}^2}{\k} \right)  2  \Re\bigg(\bigg(   \I \left( - t+ {\col f_z}  + \frac{2 p_z}{p} \right)\left (  e^{-\I(\wt r+ s)}+1 \right )   - \I t  + s_z \bigg) d z\bigg) 
\end{multline*}
simplifies  into 
\begin{multline} \label{last}
   \bigg(3 t_2\k
    -f_y\k 
    +\frac{2p_y\k}{p}  
      -2r' t_1
      -f_xr'
      +\frac{2 p_x r'}{p}     
  + \frac{r'{}^2}{\k}t_2 
   -\frac{   \k_x r'}{\k}  \bigg)dx+\\
     +\bigg( 3t_1\k
     + f_x\k  
      -\frac{2p_x \k}{p }   
       +2r' t_2  
       -f_yr'   
        +\frac{2 p_y r'}{p}    
  +\frac{r'{}^2}{\k} t_1  
   -\frac{   \k_y r'}{\k}   \bigg)dy.   
\end{multline}
Since 
$\k=  2\sign(g) p^2 e^{-f}$,   $\frac{2p_x \k}{p}=\k_x +\k f_x$ and  $\frac{2p_y \k}{p}=\k_y +\k f_y$, \eqref{last} is equal to  
\beq \label{last1} \k_y d x - \k_x d y + 
   \bigg(3 t_2\k
      -2r' t_1 
  + \frac{r'{}^2}{\k}t_2 
   \bigg)dx
     +\bigg( 3t_1\k 
       +2r' t_2  
  +\frac{r'{}^2}{\k} t_1     \bigg)dy  \ .
\eeq
 From \eqref{A5} and \eqref{last1}, we get that $g$ is as in   \eqref{111}.
It  only remains to show that $ \cH$ is equal to the sum of the three functions $h_o$, $h_m$ and $h_t$. But this is  a straightforward consequence of   \eqref{33}  and the definition \eqref{bigchange} of $\k$ and $r'$.


\begin{thebibliography}{21}
 \bibitem{ACHK}
 D.\ V.\  Alekseevsky, 
  V.\ Cort{\'e}s,
  K.\ Hasegawa and
  Y.\ Kamishima,
 {\it Homogeneous locally conformally K\"ahler and Sasaki manifolds},
   Internat. J. Math.,
   {\bf 26},
   (2015),
 1541001, 29.
   
  \bibitem{AGS} D.\ V.\  Alekseevsky,  M.\ Ganji and G.\ Schmalz,
{\it CR-geometry and shearfree Lorentzian geometry}, 
in  in ``Complex geometric analysis'', Springer Proc. Math. Stat. vol. 246, pp. 11--22, {\it Springer, Singapore}, 2018.


\bibitem{AGSS} D.\ V.\ Alekseevsky,
M.\ Ganji,
G.\ Schmalz and A.\ Spiro,
{\it Lorentzian manifolds with shearfree congruences and
 K\"{a}hler-Sasaki geometry},
Differential Geom. Appl.,
{\bf 75} (2021), Paper No. 101724, 32
%
\bibitem{AGSS1} D.\ V.\ Alekseevsky, M.\ Ganji, G.\ Schmalz and A.\ Spiro, {\it The Levi-Civita connections of Lorentzian manifolds with prescribed optical geometries}, CUBO, A Mathematical Journal, {\bf 26} (2024), 239–258.
%
\bibitem{BRT}
M.\ S.\ Baouendi, L.\ P.\ Rothschild, and F.\ Tr\`eves,
{\it CR structures with group action and extendability of CR functions},
Invent. Math. {\bf 82} (1985),  359--396.
%
\bibitem{CH} R.\ Courant and D.\ Hilbert, Methods of Mathematical Physics, Vol. II, {\it Interscience Publ., New York,} 1962.
%
\bibitem{EKS1}
V.\ Ezhov, M.\ Kol\'{a}\v{r}, and G.\ Schmalz,
{\it Rigid embeddings of Sasakian hyperquadrics in $\mathbb{C}^{n+1}$},
J. Geom. Anal. {\bf 28} (2018),  2185--2205.
%
\bibitem{FLT} A.\ Fino, T.\ Leistner and A.\ Taghavi-Chabert, {\it Optical geometries}, Ann. Sc. Norm. Super. Pisa Cl. Sci. (5), {\bf  XXVI} (2025), 341-396.
%
\bibitem{GGSS} 
M.\ Ganji, C.\ Giannotti,
G.\ Schmalz and A.\ Spiro,
{\it Einstein manifolds  with  optical geometries of Kerr type},
Ann. Physics 
{\bf 75} (2025), Paper No. 169908, 28.
%
\bibitem{GSS} M.\ Ganji, G.\ Schmalz and D.\ Sykes, {\it Quasi-{E}instein shearfree spacetimes lifted from {S}asakian
              manifolds} in ``Complex geometric analysis'', Springer Proc. Math. Stat. vol. 481, pp. 69--86, {\it Springer, Singapore}, 2025.
 %
 \bibitem{GT} D. Gilbarg and N. S. Trudinger, Elliptic Partial Differential Equations of Second Order, {\it Springer-Verlag, Berlin}, 2001.
%
\bibitem{GHN} A.\ R.\ Gover, C.\ D.\ Hill and  P.\ Nurowski, {\it Sharp version of the Goldberg-Sachs theorem}, Ann. Mat. Pura Appl. (4),  {\bf 190} (2011),  295--340. 
%
\bibitem{HLN} C.\ D.\ Hill, J.\ Lewandowski and P.\ Nurowski, 
{\it Einstein's equations and the embedding of 3-dimensional CR
   manifolds},  Indiana Univ. Math. J.
  {\bf 57}  (2008),  3131--3176.
  %
  \bibitem{Jac}
H.\ Jacobowitz,
{\it The canonical bundle and realizable CR hypersurfaces},
Pacific J. Math. {\bf 127} (1987),  91--101.
%
\bibitem{ONeill} B.\ O'Neill, The geometry of Kerr black holes, {\it A K Peters, Ltd., Wellesley, MA}, 1995.
%
      %
 \bibitem{NT} {\col P. Nurowski and J. Tafel, 
{\it New algebraically special solutions of the
              {E}instein-{M}axwell equations},
Classical Quantum Gravity
 {\bf 9} (1992), 2069--2077.}
 %
\bibitem{Rob} I.\ Robinson, {\it Null electromagnetic fields}, J. Mathematical Phys. {\bf 2}
(1961), 290--291.
%
\bibitem{RT} I.\ Robinson and A.\ Trautman, {\it Conformal geometry of flows in {$n$} dimensions},
J. Math. Phys.  {\bf 24} (1983), 1425--1429. 
%
\bibitem{RT11}  I.\ Robinson and A.\ Trautman,
{\it Cauchy-{R}iemann structures in optical geometry},
in  ``Proceedings of the Fourth {M}arcel {G}rossmann meeting on
              general relativity, {P}art {A}, {B} ({R}ome, 1985)'',
pp. 317--324,
 {\it North-Holland, Amsterdam},
1986.
%
\bibitem{RT1} I.\ Robinson and A.\ Trautman, {\it Optical geometry} in 
``Warsaw symposium on elementary particle physics: new theories in physics,
Kazimierz (Poland), 1988'', pp. 454--497, 
 {\it World Scientific Pub. Co., Teaneck, NJ (USA)},
1989. 
%
\bibitem{SKH} A.\ A.\ Shaikh, Y.-H.\ Kim,  and S.\ K.\ Hui,
 {\it On {L}orentzian quasi-{E}instein manifolds},
J. Korean Math. Soc.  {\bf 48}
     (2011), 669--689.
%
\bibitem{Tafel1} J.\ Tafel, {\it On the {R}obinson theorem and shearfree geodesic null
              congruences}, Lett. Math. Phys. {\bf 10} (1985),  33--39. 
%
\bibitem{Traut3} A.\ Trautman, {\it Robinson manifolds and the shear-free condition}, 
in  {``Proceedings of the {C}onference on {G}eneral {R}elativity,
              {C}osmology and {R}elativistic {A}strophysics ({J}ourn\'ees
              {R}elativistes) ({D}ublin, 2001)}''
Internat. J. Modern Phys. A {\bf 17} (2002),  2735--2737.
\end{thebibliography}
\end{document}